\documentclass[12pt,letterpaper,notitlepage]{article}

\usepackage{epsfig}
\usepackage{amsmath,amssymb}
\usepackage{cite}
\usepackage{color}
\usepackage{subfigure}
\usepackage{axodraw4j}
\usepackage{ulem}

\newcommand{\tev}{\,\, \mathrm{TeV}}
\newcommand{\gev}{\,\, \mathrm{GeV}}

\newcommand{\lesim}{\,\raisebox{-.1ex}{$_{\textstyle<}\atop^{\textstyle\sim}$}\,}
\newcommand{\gesim}{\,\raisebox{-.1ex}{$_{\textstyle>}\atop^{\textstyle\sim}$}\,}
\newcommand{\h}{H^0}
\begin{document}

\begin{titlepage}

\begin{flushright}
PITT-PACC-1405
\end{flushright}

\vspace{15pt}
\begin{center}
\LARGE {Top-Quark Initiated Processes \\ at High-Energy Hadron Colliders}
\end{center}

\vspace{0pt}
\begin{center}
{\large Tao Han, Josh Sayre, and Susanne Westhoff}\\
\vspace{30pt} {
PITTsburgh Particle-physics Astro-physics \& Cosmology
    Center (PITT-PACC),\newline  Department of Physics \& Astronomy,
    University of Pittsburgh, Pittsburgh, PA 15260, USA
   }
\end{center}

\vspace{10pt}
\begin{abstract}
\vspace{2pt} 
\noindent
In hadronic collisions at high energies, the top-quark may be treated as a parton inside a hadron. Top-quark initiated processes become increasingly important since the top-quark luminosity can reach a few percent of the bottom-quark {lu\-mi\-nosi\-ty}. In the production of a heavy particle $H$ with mass $m_H > m_t$, treating the top-quark as a parton allows us to resum large logarithms $\log(m_{H}^{2}/m_{t}^{2}$) arising from collinear splitting in the initial state.
We quantify the effect of collinear resummation at the 14-TeV LHC and a future 100-TeV hadron collider, focusing on the top-quark open-flavor process $gg\to t\bar t H$ in comparison with $t\bar t \to H$ and $tg\rightarrow tH$ at the leading order (LO) in QCD. We employ top-quark parton distribution functions with appropriate collinear subtraction and power counting.
We find that (1) Collinear resummation enhances the inclusive production of a heavy particle with $m_H^{}\approx 5\tev \ (0.5\tev)$ by more than a factor of two compared to the open-flavor process at a 100-TeV (14-TeV) collider; (2) Top-quark mass effects are important for scales $m_H$ near the top-quark threshold, where the cross section is largest. We advocate a modification of the ACOT factorization scheme, dubbed m-ACOT, that consistently treats heavy-quark masses in hadronic collisions with two initial heavy quarks; (3) The scale uncertainty of the total cross section in m-ACOT is of about $20\%$ at the LO. While a higher-order calculation is indispensable for a precise prediction, the LO cross section is well described by the process $t\bar t\to H$ using an effective factorization scale significantly lower than $m_H$.
We illustrate our results by the example of a heavy spin-0 particle. Our main results also apply to the production of particles with spin-1 and 2.
\end{abstract}

\end{titlepage}

\clearpage

\section{Introduction}
The milestone discovery of the Higgs boson completes the Standard Model (SM) of particle physics for the electroweak and strong interactions. Although, with a light Higgs boson, the theory can be valid all the way up to the Planck scale, there are theoretical and observational indications of the need for new physics beyond the SM. The search for new physics at the TeV scale continues. One obvious route is to study the Higgs properties to a high precision at the LHC and a future Higgs factory \cite{SnowHiggs} to seek for deviations from the SM predictions. The other approach is to push the energy frontier beyond the LHC regime in the hope of reaching a new physical threshold \cite{Gershtein:2013iqa}.

The perspective of a proton-proton collider at the ultra-high center-of-mass (CM) energy of $\sqrt{S}=100\tev$ (VLHC) would lead us to a new territory far beyond the reach of the LHC. We hope to produce new heavy particles associated with a new sector beyond the SM. On the other hand, all particles in the SM would appear essentially massless in processes far above the electroweak scale, in the unbroken phase of the electroweak sector. This feature implies new phenomena even within the SM, such as electroweak bremsstrahlung \cite{Hook:2014rka}, a strong enhancement from heavy quarks produced collinear to the beam line \cite{Barnett:1987jw,Olness:1987ep}, and highly boosted objects from the decay of the electroweak gauge bosons, the Higgs boson and the top-quark \cite{Abdesselam:2010pt}. As the heaviest SM particle, the top-quark may hold the promise to be sensitive to a new-physics sector. Yet, at a CM energy of $100\tev$, the top-quark would be as ``massless'' as the bottom-quark at the Tevatron. This motivates us to study the characteristic behavior of the top-quark in the ultra-high energy regime in QCD, as well as its role in searching for new physics.

Collinear enhancement occurs in top-quark initiated processes involving an energy scale $Q$ above the top mass $m_t$, manifesting itself as a logarithmic factor 
$\alpha_s \log(Q^2/m_t^2)$ in the perturbative cross section, where $\alpha_s$ is the coupling constant of QCD. At a sufficiently large scale $Q$, where the collinear region of the phase space dominates the total cross section, it is justified to consider the top-quark as an active parton. The definition of a top-quark parton distribution function (PDF) inside the proton allows one to resum collinear logarithms to all orders $\alpha_s^n \log^n(Q^2/m_t^2)$ via DGLAP evolution \cite{Altarelli:1977zs}. In this limit, the leading partonic process for the production of a heavy state $H$ with mass $m_{H} \gg m_{t}$ is top-quark fusion $t\bar t\rightarrow H$. The gluon-induced processes $g t\rightarrow t H$ and $gg\rightarrow t\bar t H$ enter as sub-leading corrections of the order of $\alpha_s$ and $\alpha_s^2$, respectively, once the collinear region is appropriately subtracted. However, care needs to be taken when the new scale is not far above the top-quark mass, $m_H\gesim m_t$.

Factorization schemes with a variable number of active quark partons \cite{Barnett:1987jw,Olness:1987ep,Thorne:1997ga} interpolate between the $N$-flavor scheme with a massive heavy quark near its production threshold and the $N+1$-flavor scheme with massless quarks at high scales. The implementation of such a scheme requires a subtraction mechanism to avoid double-counting in the collinear region when including resummation effects.
The early work by Aivazis, Collins, Olness and Tung (ACOT) \cite{Aivazis:1993pi} laid out a consistent scheme in which initial heavy quarks are set on-shell and the quark mass is kept throughout in the partonic cross section. Collins later demonstrated by the example of deeply inelastic scattering (DIS) that factorization holds if quarks are treated massless in partonic sub-processes with initial heavy quarks, but are kept massive in gluon-initiated processes (a scheme known as simplified ACOT or s-ACOT) \cite{Collins:1998rz,Kramer:2000hn}. While neglecting the quark mass entirely is in general not justified, especially for scales near the heavy-quark threshold, the s-ACOT scheme is applicable for arbitrary scales above the quark threshold in DIS.

However, when applied to a process with two initial heavy quarks, care needs to be taken with  the mass treatment in s-ACOT. We study in detail the effects of the heavy-quark mass near the top-quark threshold and in the intermediate energy region. We propose an appropriate factorization prescription and provide one single scheme (to be called modified ACOT scheme, or m-ACOT scheme), which allows a correct description of the production of a massive resonance from heavy-quark fusion in a broad range of scales $Q\gtrsim m_t$.
We furthermore examine the dependence of physical observables on the factorization scale. In this context, we estimate the ``effective scale'' of collinear gluon splitting into heavy quarks, which lies far below the relevant high-energy scale $Q$ in the process.

With respect to new-physics searches, we investigate the production of heavy particles with spin-0, 1 and 2 from heavy-quark fusion at tree level and discuss the respective behavior for $m_H$ near the top threshold. For the sake of illustration, we present our results by the example of a heavy spin-0 particle with a scalar coupling to top-quarks. Our main results on factorization and resummation are generally applicable to the other cases.

Heavy-quark initiated processes have been studied in the literature, including neutral Higgs boson production from bottom-quark fusion $b\bar b\to \h$ \cite{Dicus:1988cx,Dicus:1998hs,Maltoni:2003pn,Maltoni:2012pa}, and charged-Higgs production via $t\bar b \to H^+$ \cite{Barnett:1987jw,Olness:1987ep,Dawson:2014pea}. We will compare our results with some of these works in Sections~\ref{sec:massless} and \ref{sec:dis}.

The rest of the paper is organized as follows. In Section~\ref{sec:proc}, we lay out the formalism for top-quark initiated processes. We discuss the parton luminosities including top-quarks at the 14-TeV LHC and at a 100-TeV circular proton-proton collider. We also give the squared matrix elements for the production of particles with spin-0, 1, and 2 from heavy-quark fusion. In Section~\ref{sec:schemes}, we briefly review the factorization schemes ACOT, s-ACOT and the massless quark limit for treating heavy quarks as partons and advocate a modified scheme (m-ACOT) for a consistent treatment of processes with two initial heavy quarks. In Section~\ref{sec:vlhc}, we investigate the dependence of physical observables on the factorization scale and derive an ``effective scale'' to optimize fixed-order calculations in the 6-flavor scheme. We discuss the extension of our work beyond the LO and compare our results with the literature in Section~\ref{sec:dis} and conclude in Section~\ref{sec:sum}. In Appendix~\ref{app}, we give a detailed and more general discussion of factorization and quark mass effects. Appendix~\ref{sec:part-xs} contains analytic formulae for the partonic cross section of neutral scalar production from top-quark fusion.


\section{Top-quark initiated processes}
\label{sec:proc}
Effects of initial top-quarks are generally important in processes that are sensitive to a high energy scale, such as the production of a very heavy particle or the high-energy regime in differential distributions. In particular, certain new particles may preferably couple to heavy quarks. Well-known examples include the Higgs boson(s) \cite{Dittmaier:2011ti}, new vector bosons in models with strong dynamics \cite{Hill:2002ap}, and Kaluza-Klein gravitons \cite{KK}.
We first parameterize the generic couplings of the heavy particles with spin 0, 1 or 2 to heavy quarks as
\begin{eqnarray}
\label{eq:coupling}
&& {\rm spin\ 0:\ \ neutral\ scalar}\ H^0: \ \  i {y \over \sqrt 2};\quad
{\rm pseudo\ scalar}\ A^0:\ \  i {y \over \sqrt 2}\gamma_{5};\\\nonumber
&& \hspace*{1.7cm}{\rm charged\ scalar}\ H^+:\ \  i {y \over \sqrt 2}(g_{L} P_{L} + g_{R} P_{R} ) ; \\\nonumber
&& {\rm spin\ 1:\ \ color-singlet\ vector/axial\ vector}\ Z^{'0},\,W^{'+}:\ \   i g \gamma^{\mu}(g_V^{}-  g_A^{} \gamma_{5} ); \\\nonumber
&& \hspace*{1.7cm}{\rm  color-octet\ vector/axial\ vector}\ g^{}_{KK}:\ \ i g_s \gamma^{\mu}(g_V^{}-  g_A^{} \gamma_{5} )\,t^a; \\\nonumber
&& {\rm spin\ 2:\ \ tensor}\ G:\ \   -i {\kappa \over 8} [ 
\gamma^\mu (p_t - p_{\bar t})^\nu + 
\gamma^\nu (p_t - p_{\bar t})^\mu 
-2g^{\mu\nu}(\slash{\hskip -0.2cm}p_t - \slash{\hskip -0.2cm}p_{\bar t} -2m_t)
].
\end{eqnarray}
  For the production of a charged boson $H^\pm,\,W^{'\pm}$ from fermions with different masses, there appears an extra kinematic factor $\ell$, which modifies the threshold behavior with respect to pure S-wave and P-wave production in the case of equal fermion masses.

\begin{table}
\caption{Spin- and color-averaged squared matrix elements for the production of an on-shell heavy particle of mass $m_H=\sqrt{s}$ from heavy-quark fusion and the corresponding threshold behavior. The number of colors is denoted by $N_c$ and the $\rm{SU}(3)$ invariant as $C_F=4/3$. Subscripts $T$ and $L$ indicate transverse and longitudinal polarization, respectively. The kinematic factors are $\beta^{2}_{ij} = \ell_{ij}(1-(m_i+m_j)^2/s)$ and $\ell_{ij} = 1-(m_i-m_j)^2/s$, as well as the couplings $g_{S,P} = (g_{L}^{} \pm g_{R}^{})/2$ in terms of chiral couplings $g_L$ and $g_R$.}
 \begin{center}
\begin{tabular}{|c|c|c|}
\hline 
process & $\overline\sum |{\cal M}|^{2}$ & threshold behavior
\tabularnewline\hline\hline
$t\bar t \to H^0$ & {\large{${y^2 s \over 4N_{c}}$}} $\beta_{t\bar t}^{2}$ & P-wave \tabularnewline\hline
$t\bar t \to A^0$ & {\large{$ {y^2 s \over 4N_{c}} $}} &  S-wave \tabularnewline\hline 
$t\bar b \to H^+$ & {\large{${y^2 s \over 4N_{c}}$}} $(g_{S}^{2} \beta_{t\bar b}^{2} /\ell_{t\bar b}  + g_{P}^{2} \ell_{t\bar b} )$ &
same as $H^0$, $A^0$, with an extra $\ell$ \tabularnewline\hline\hline
$t\bar t \to Z^{'0}_{T}$ & {\large{${g^2 s \over N_{c}}$}} $( g_{V}^{2} + g_{A}^{2}\beta_{t\bar t}^{2} )$ & vector:\,S-wave;\,axial-vector:\,P-wave \tabularnewline
$t\bar t \to Z^{'0}_{L}$ & {\large{${g^2 s \over N_{c}}$}} $ g_{V}^{2} (2m_{t}^{2}/s) $ & fermion mass suppression
\tabularnewline\hline 
$t\bar b \to W^{'+}_{T}$ & {\large{${g^2 s \over N_{c}}$}} $\big( g_{V}^{2} \ell_{t\bar b} + g_{A}^{2} \beta_{t\bar b}^{2}/\ell_{t\bar b} \big)$ & same as $Z^{'0}_T$, with an extra $\ell$ \tabularnewline 
$t\bar b \to W^{'+}_{L}$ & {\large{${g^2 s \over N_{c}}$}} $\big( g_{V}^{2} \ell_{t\bar b}$ {\large{$ { (m_{t}+m_{b})^{2}\over 2s} $}} + $g_{A}^{2} \beta_{t\bar b}^{2}$ {\large{$ {(m_{t}-m_{b})^{2}\over 2s \ell_{t\bar b}}$}}\big) & fermion mass suppression \tabularnewline\hline
$t\bar t \to g^{}_{KK}$ & $C_F${\large{${g_s^2 s \over N_{c}}$}} $( g_{V}^{2}(1+2m_t^2/s) + g_{A}^{2}\beta_{t\bar t}^{2} )$ & same as $Z^{'0}$
\tabularnewline\hline\hline
$t\bar t \to G$ &  {\large${\kappa^2 s^{2} \over 16 N_{c}}$} $(1+8m_{t}^{2}/3s) \beta_{t\bar t}^{2} $ & P-wave \tabularnewline\hline
\end{tabular}
\end{center}
\label{tab:sqme}
\end{table}
According to the QCD factorization theorem, the total inclusive cross section for the hadronic production of a heavy particle $H$ can be expressed as 
\begin{align}\label{eq:factorization}
\sigma_{pp\rightarrow H+X}(S) & =  \sum_{i,j} \int_{m_H^2/S}^1 dx_1\int_{m_H^2/(x_1S)}^1 dx_2\, f_{i}(x_1,\mu)\,f_{j}(x_2,\mu)\,\hat{\sigma}_{i j\rightarrow H}(s)\\\nonumber
& \equiv 
\sum_{i,j} \int_{m_H^2/S}^{1} d\tau\ {dL_{ij}\over d\tau}\ \hat{\sigma}_{ij}(s),\qquad {dL_{ij}\over d\tau}(\tau,\mu) = \int_\tau^1\frac{\text{d}x}{x}f_i(x,\mu)f_j(\tau/x,\mu),
\end{align}
where $f_{i,j}(x,\mu)$ are the PDFs of partons $i,j=\{q,\bar q,g\}$ with momentum fraction $x$ inside the proton, $\mu$ denotes the factorization scale, $\sqrt{s}$ and $\sqrt{S}$ are the partonic and hadronic CM energies, and $\tau \equiv s/S = x_{1} x_{2}$.  

We assume that heavy quarks inside the proton are dynamically generated by QCD interactions. Therefore we set the heavy-quark PDFs to zero for scales below the quark mass and evolve them to higher scales by including them in the DGLAP equations, beginning at the mass threshold. We have carried out this evolution to LO for the top-quark PDF $f_t(x,\mu)$ numerically. As an input for the gluon and light-quark PDFs at the initial scale $\mu = m_t$ we use the NNPDF2.3 LO distributions of the NNPDF collaboration \cite{Ball:2012cx}. The NNPDF collaboration has released top-quark PDFs as part of the NNPDF2.3 set up to NNLO in the range $1.4\,\rm{GeV}<\mu<10\,\rm{TeV}$. Analytically, our top-quark PDF corresponds with the NNPDF2.3 version at LO. We have checked this agreement numerically. NNPDFs are derived in the so-called FONLL scheme, which is equivalent to the s-ACOT scheme up to (at least) NLO~\cite{Forte:2010ta}. For practical purposes, NNPDF2.3 top-quark PDFs can thus be used for calculations in the ACOT scheme and its variations s-ACOT and m-ACOT. In our numerical analysis, we set the factorization scale and the renormalization scale equal and, unless stated otherwise, fixed to the heavy particle mass 
\begin{equation}
\mu = m_H. 
\label{eq:mu}
\end{equation}

\begin{figure}[t]
\begin{centering}
\begin{tabular}{cc}
\hspace*{-0.2cm}\raisebox{4.6cm}{(a)}\hspace*{-0.2cm}\includegraphics[width=7.5cm]{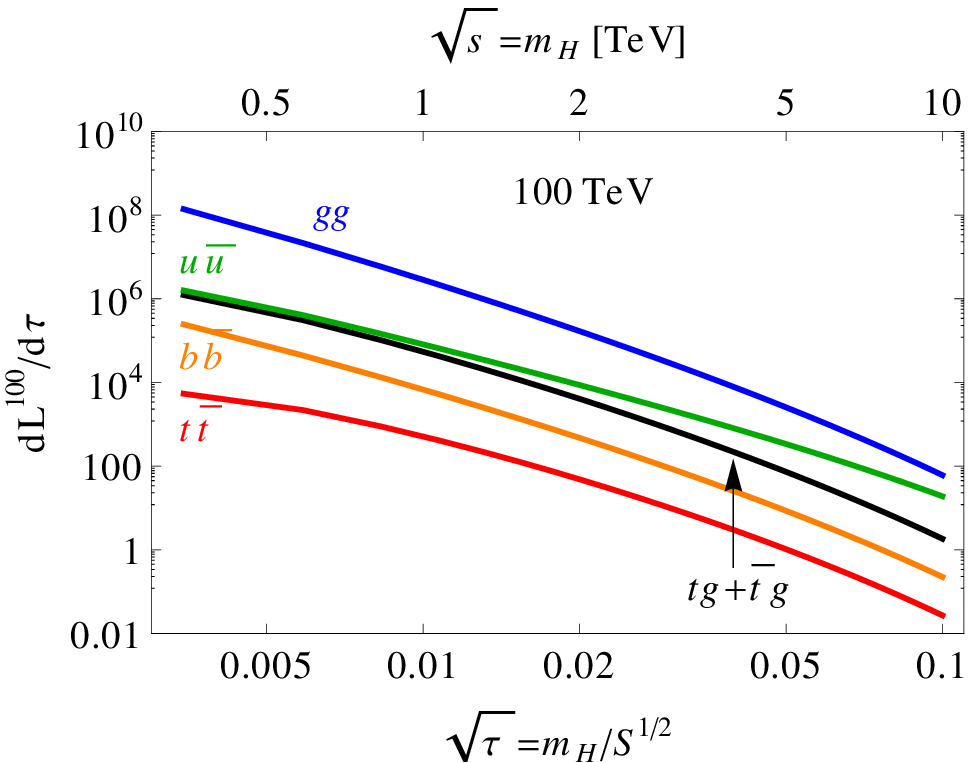} &
\hspace*{0.2cm}\raisebox{4.6cm}{(b)}\hspace*{-0.2cm}\includegraphics[width=7.5cm]{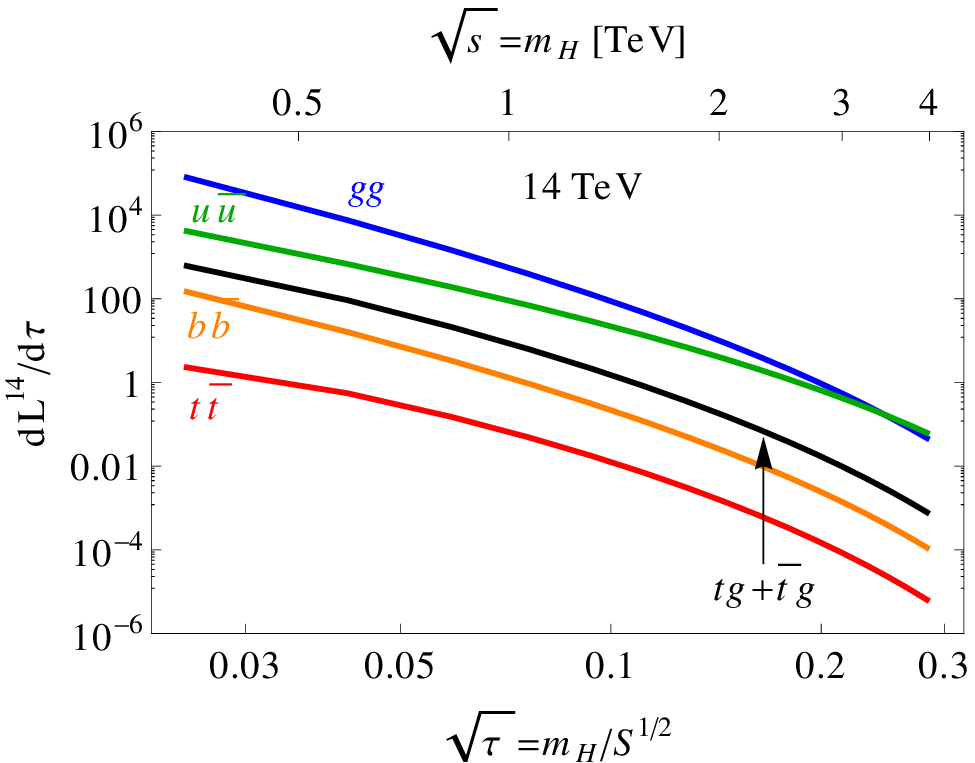}
\end{tabular}\vspace*{0.3cm}
\raisebox{4.5cm}{(c)}\hspace*{-0.2cm}\includegraphics[width=7.5cm]{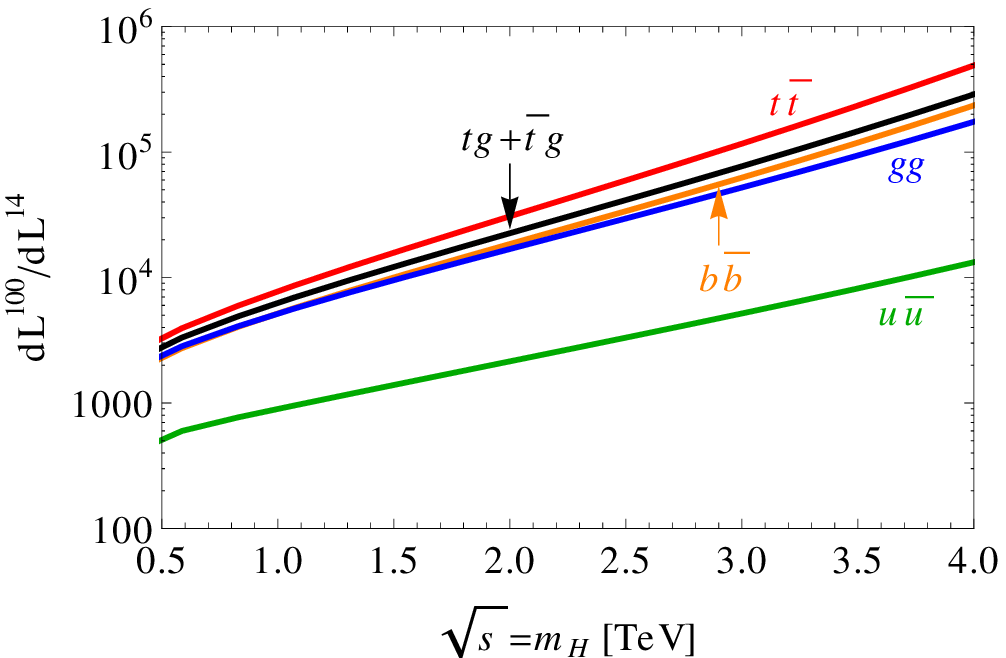}
\caption{Parton luminosities (a) at $\sqrt{S}=100\tev$, (b) at $\sqrt{S}=14\tev$, and (c) ratio between 
those at 100 TeV and 14 TeV, for $t\bar t$ (red), $tg+\bar tg$ (black), $b\bar b$ (orange), $u\bar u$ (green) (including the initial-state interchange), and $gg$ (blue).
}
\label{fig:lumi}
\end{centering}
\end{figure}

An estimate of the relevance of initial top-quarks in high-energetic processes can be obtained by considering the parton luminosities $dL_{ij}/d\tau$, which depend only on $\tau$ and $\mu$. 
 Figures~\ref{fig:lumi}(a) and \ref{fig:lumi}(b) show parton luminosities at $\sqrt{S}=100$ and $14\tev$, respectively, with top-quarks in comparison to light quarks and gluons. We present them as functions of $\sqrt{\tau} = m_{H}/\sqrt{S}$, which indicates the geometrical mean of the energy fractions, $\bar x \equiv \sqrt{x_1 x_2} = \sqrt{\tau}$, in the resonant production of $H$. The range of the partonic CM energy $\sqrt{s}=m_H$, labelled on the top axis, extends from the top-quark threshold up to $\sqrt{s}=10\tev$ ($4\tev$) at $\sqrt{S}=100\tev$ ($14\tev$), which corresponds to a $t\bar t$ luminosity of about $0.01\ (10^{-5})$. This defines the kinematic range of our current interest at the VLHC,
\begin{equation}
 0.002  \lesim \bar x \lesim  0.1, \quad {\rm for}\quad 200\gev \lesim \sqrt s \lesim 10\tev.
\end{equation}
 We see that the gluon-gluon ($gg$) luminosity (blue, top curve) is overwhelmingly dominant, exceeding the top-antitop ($t\bar t$) luminosity (red, bottom curve) by three to four orders of magnitude in this range. For comparison, we also show the luminosities for bottom-quarks (orange) and up-quarks (green). At these energies, the top-quark luminosity $L_{t\bar t}$ is only about an order of magnitude smaller than $L_{b\bar b}$. The valence-quark luminosity $L_{u\bar u}$ becomes substantial for $\bar x \gesim 0.1$. 

To put this into perspective, we compare in Figure~\ref{fig:lumi}(c) the luminosity ratio between $\sqrt{S}=100\tev$ and $\sqrt{S}=14\tev$, $dL^{100}/dL^{14}$ in the relevant energy range at the LHC, $200\gev \lesssim \sqrt s \lesssim 4\tev$. We see that the parton luminosities for $t\bar t$, $tg$, $b\bar b$, and $gg$ increase by a large factor of about $10^3 - 10^5$. The luminosity enhancement for valence quarks $u\bar u$ is more modest, ranging between $500$ and $10^4$. We note that the luminosity enhancement for top- and bottom-quarks is comparable to that for $gg$ close to the top-quark threshold, but exceeds it significantly at higher scales $\mu=m_H$. This is due to the stronger suppression of heavy-quark PDFs versus gluon PDFs at high $x$, which dominates the luminosity at high scales at the LHC.

\begin{figure}[!t]
\begin{centering}
\begin{tabular}{cc}
\includegraphics[width=8cm]{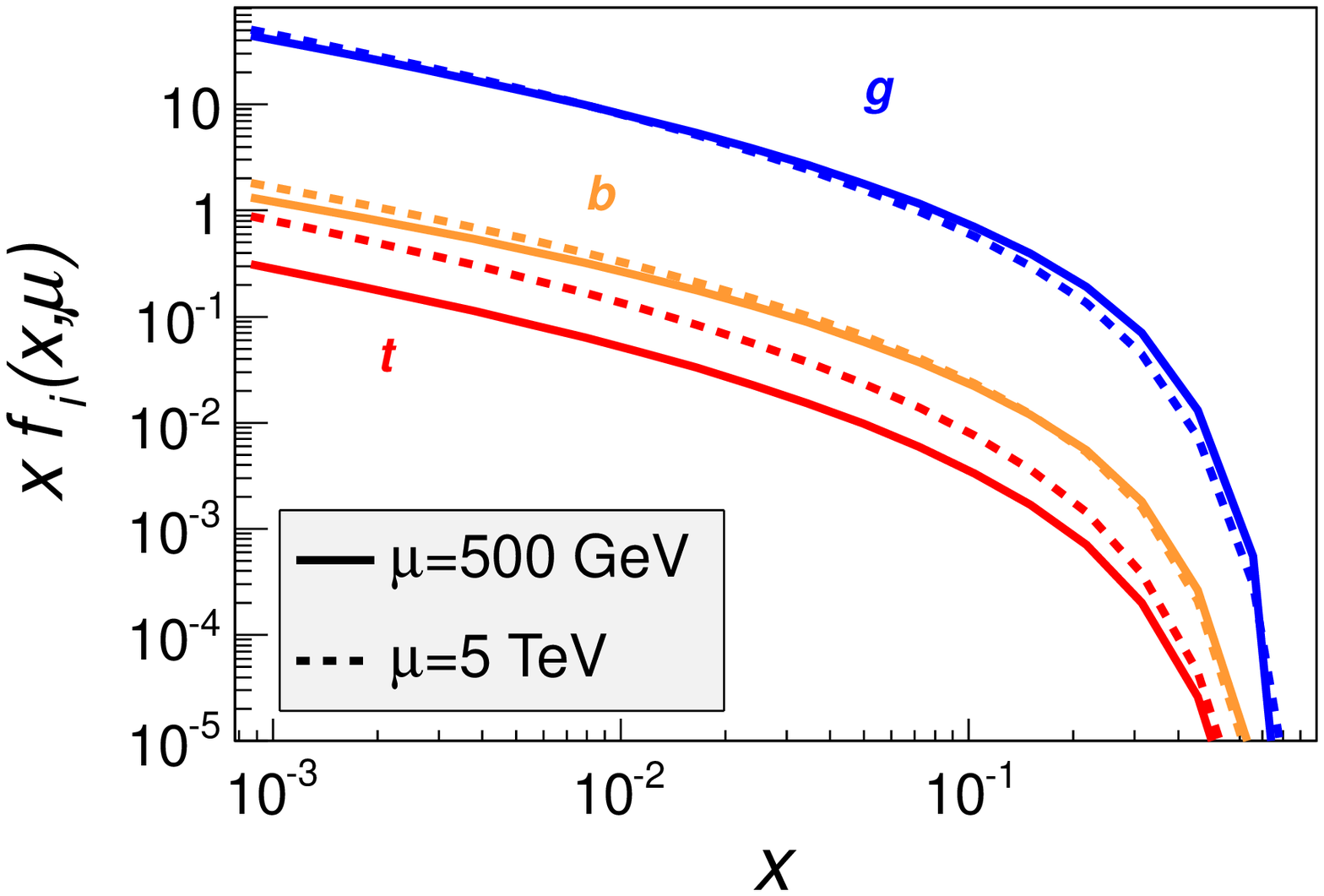}
\includegraphics[width=8cm]{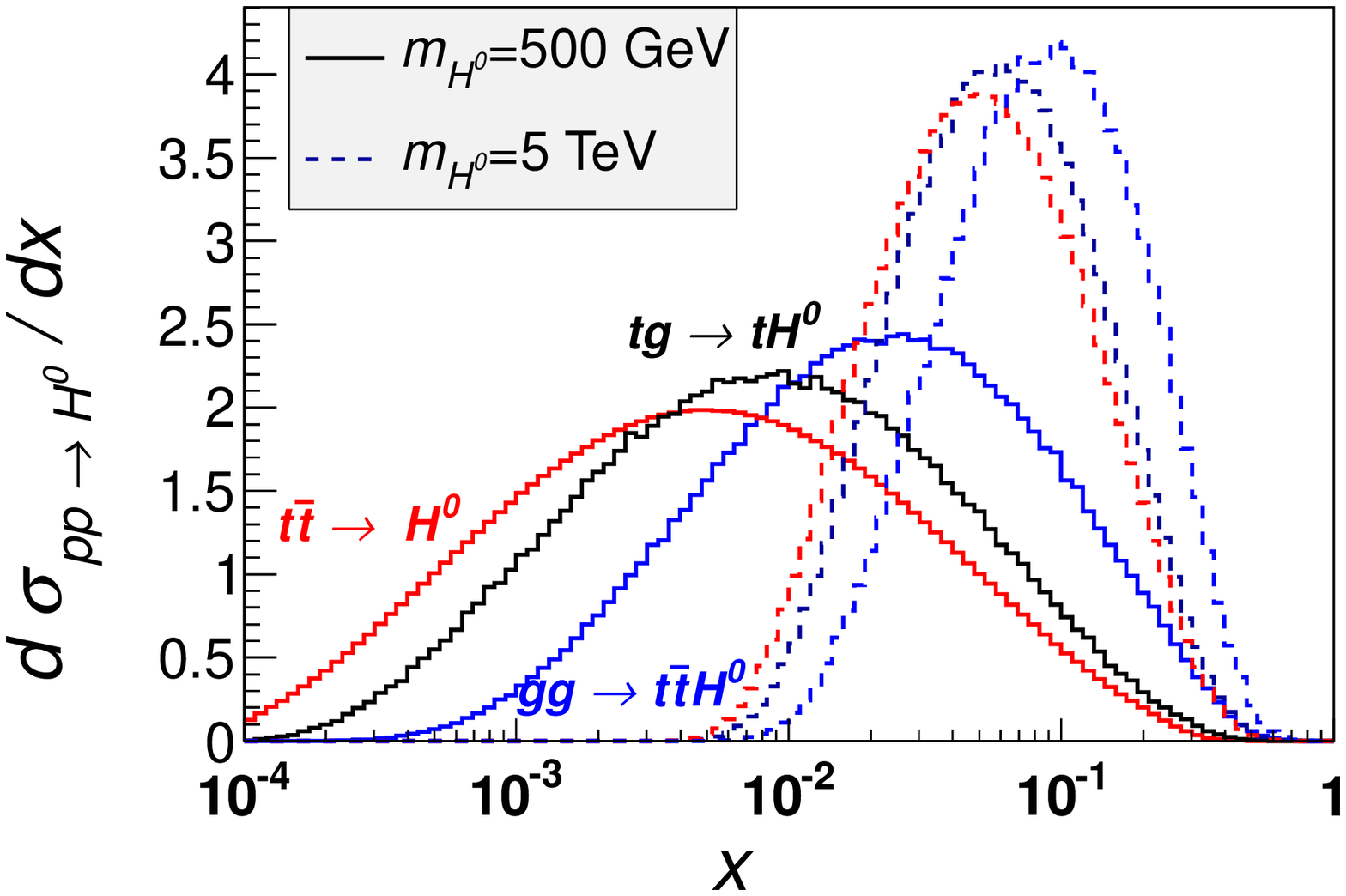}
\end{tabular}
\caption{Parton distributions at $\sqrt{S}=100\tev$. Left: momentum distribution $x f(x,\mu)$ versus $x$ for the top-quark (red), bottom-quark (orange), and gluon (blue) at $\mu=500\gev$ (solid) and $\mu = 5\tev$ (dashed). Right: $x$ distribution of the cross section for resonant scalar production in Eq.~(\ref{eq:factorization}) from the partonic processes $t\bar t\to H^0$ (red), $tg\to tH^0$ (black), and $gg\to t\bar t H^0$ (blue) for $m_{H^0}=500\gev$ (solid) and $m_{H^0}= 5\tev$ (dashed).
}
\label{fig:pdfx}
\end{centering}
\end{figure}

In Figure~\ref{fig:pdfx}, we present the individual parton distributions at $100\tev$. Figure~\ref{fig:pdfx}, left panel, shows the parton momentum distribution $x f(x,\mu)$ versus $x$ for the top-quark (red), bottom-quark (orange), and gluon (blue) at $\mu=500\gev$ (solid) and $\mu=5\tev$ (dashed). While the general tendency is similar to the luminosities in Figure~\ref{fig:lumi}(a), we see explicitly that the top-quark PDF is less than an order of magnitude lower than the bottom-quark PDF at $\mu = 500\gev$ and grows in relative importance at higher scales. Figure~\ref{fig:pdfx}, right panel, shows the differential cross section in Eq.~(\ref{eq:factorization}) for the production of a neutral scalar $H^0$ versus the momentum fraction $x_1$ of the first parton
for representative processes $t\bar t \to H^0$ (red), $tg \to t H^0$ (black) and $gg \to t \bar t H^0$ (blue). At a low mass $m_{H^0}=500\gev$ (solid curves), a broad range of $x$ is being probed, which narrows down at higher masses, as illustrated for $m_{H^0}=5\tev$ (dashed curves). The peak value of $x_1\approx 0.005\,(0.05)$ at $m_{\h}=500\gev\,(5\tev)$ for $t\bar t\to \h$ corresponds to the average $\bar{x}=\sqrt{\tau}$ in Figure~\ref{fig:lumi}(a) and moves to higher values for the processes $tg\to t\h$ and $gg\to t\bar t\h$.

Overall, simply based on these parton luminosity arguments, we see that top-quark initiated processes become increasingly important 
\begin{itemize}
\item[(i)] for a fixed CM energy $\sqrt{S}$: at higher scales $\sqrt{s}\approx \mu \approx m_H$ due to the large collinear logarithms; 
\item[(ii)] for a fixed scale $m_H$: at higher CM energies $\sqrt{S}$ due to the higher parton density at smaller $\bar x$.  
\end{itemize}
The relative importance for top-quark initiated processes may, however, go well beyond the PDF enhancement. They become especially relevant in models with flavor-hierarchical couplings that are stronger for the top-quark. Furthermore, an initial heavy quark generally induces a partonic process of lower order in $\alpha_s$ and with fewer particles in the final state than an initial gluon. 

\section{Factorization schemes and top-mass effects}\label{sec:schemes}
The validity of factorization and the use of top-quark PDFs strongly depend on the different energy scales involved in the process.
\begin{itemize}
\item At  energy scales below and around the top mass, $m_H\lesssim m_t$, it is appropriate to work with five active quark flavors and not consider the top-quark as a constituent of the proton. The production of a heavy particle $H$ in association with top-quarks is then reliably described at LO by the partonic process $gg\rightarrow t\bar t H$ with massive top-quarks, as depicted in Figure~\ref{fig:graphs}(a). We will refer to this case as the 5-flavor scheme.
This is the case for the production of the SM Higgs boson in association with top-quarks at the LHC \cite{Dittmaier:2011ti}.
\item When there exists a higher energy scale involving a heavy final state $m_H\gg m_t$, 
the top-quark can be considered essentially massless and an active quark flavor inside the proton. In this case, the total cross section in Eq.~(\ref{eq:factorization}) is dominated by the tree-level partonic process in the $\alpha_s$ expansion, $t\bar t\rightarrow H$, depicted in Figure~\ref{fig:graphs}(c). This approach corresponds to the 6-flavor scheme with massless quarks. The PDF treatment of top-quarks summing over the large logarithms is appropriate for a reliable prediction. The production of the SM Higgs boson from $b\bar b$ fusion serves as a good example for an analogous situation \cite{Dicus:1988cx,Dittmaier:2011ti}.
\item A non-trivial region occurs for $m_H \gesim m_t$, where corrections of $\mathcal{O}(m_t^2/m_H^2)$ can be sizeable. In this regime, the top-quark mass should not be neglected in the calculation. Since the logarithmic terms are less dominant, the contributions from higher-order matrix elements such as $tg\rightarrow t H$ and $gg\rightarrow t\bar t H$ (Figures~\ref{fig:graphs}(b) and (a)) may be equally important and thus need to be included consistently.
\end{itemize}
\begin{figure}
\begin{center}
\begin{picture}(360,80)(0,0)
\Gluon(20,0)(20,40){3}{3}
\Gluon(80,0)(80,40){3}{3}
\ArrowLine(20,40)(80,40)
\ArrowLine(20,80)(20,40)
\ArrowLine(80,40)(80,80)
\DashLine(50,40)(50,80){3}

\ArrowLine(160,0)(160,40)
\ArrowLine(160,40)(200,40)
\ArrowLine(200,40)(200,80)
\DashLine(160,40)(160,80){3}
\Gluon(200,0)(200,40){3}{3}

\Text(0,40)[t]{(a)}
\Text(140,40)[t]{(b)}
\Text(260,40)[t]{(c)}

\ArrowLine(300,40)(330,0)
\ArrowLine(270,0)(300,40)
\DashLine(300,40)(300,80){3}
\end{picture}
\end{center}
\caption{Representative tree-level graphs relevant to the inclusive production of a heavy color-singlet neutral particle (dashed line). (a) The 5-flavor LO process from $gg$ fusion; (b) The 6-flavor process involving only one initial top (solid line) in $tg$ fusion; (c) The 6-flavor process initiated by $t\bar t$ fusion.}
\label{fig:graphs}
\end{figure}
We wish to provide a clear and consistent treatment,  which is applicable  generally for the production of a heavy final state $H$ above the top-quark threshold. In this section we provide a careful analysis of top-mass effects on the cross section in the various energy regimes, focusing on the phenomenological consequences of using different factorization schemes. For a detailed and more formal discussion of mass and factorization-scale treatment for heavy quarks in general and the application to high-energy processes with collinear top-quarks, we refer the reader to Appendix~\ref{app}.

In the rest of this paper, we will focus on the production of a generic heavy neutral scalar $\h$ via top-quark fusion 
\begin{equation}
t \bar t \to \h
\end{equation}
for illustration. Analytic results for the partonic cross section of top-quark initiated neutral scalar production at LO can be found in Appendix~\ref{sec:part-xs}. We will set the top-scalar coupling in Eq.~(\ref{eq:coupling}) to $y=1\approx y_t$ for simplicity and neglect any (model-dependent) scale dependence. The general formalism below is applicable to the heavy-quark initiated production of a heavy particle with $m_{H}>m_t$ for all processes in Table~\ref{tab:sqme}. We will comment on differences with other heavy states when appropriate. 


\subsection{Massive top-quark partons: ACOT scheme}\label{sec:acot}
The ACOT scheme \cite{Aivazis:1993pi} appropriately interpolates between scales from $m_H\gtrsim m_t$ up to $m_H\gg m_t$. In this scheme the incoming top-quarks are set collinear and on-shell and in general:
\begin{center}
{\it The top-quark mass is retained throughout in the partonic cross section.}
\end{center}
The process $t\bar t \rightarrow \h$ is the first term in the ACOT expansion of the hard-scattering kernel. The top-quark PDF $f_t(x,\mu)$ gives an approximate sum of collinear parton splittings to all orders in the leading logarithm (LL), $\alpha_s^n L^n$, where $L\equiv \log(\mu^2/m_t^2)$.  However, if the new scale is not far above the top threshold, $\mu \sim m_{\h} \gesim m_t$, one needs to include potentially large contributions from 
\begin{equation}\label{eq:tgth}
 t g \to t \h, \quad gg\to t \bar t \h.
\end{equation}
In order to avoid double-counting, the collinear gluon splitting $g\to t\bar t$, which is implicit in the LL expansion of the top-quark PDF, must be subtracted. The total hadronic cross section for $\h$ production at tree level in the ACOT scheme can then be written as \cite{Dicus:1988cx,Aivazis:1993pi}
\begin{align}\label{eq:xs-acot}
\sigma_{pp\rightarrow \h} & = \Big\{[f_t-f_t^0](x_1)\times \hat{\sigma}_{t\bar t\rightarrow \h }(m_t)\times [f_{\bar t}-f_{\bar{t}}^0](x_2)\\\nonumber
&\ \ \, + [f_t-f_t^0](x_1)\times \hat{\sigma}_{tg\rightarrow t \h}(m_t)\times f_g(x_2) + (x_1\leftrightarrow x_2)\\\nonumber
&\ \ \, + \qquad\ \ f_g(x_1)\times\hat{\sigma}_{gg\rightarrow t\bar t \h}(m_t)\times f_g(x_2)\Big\} + (t\leftrightarrow \bar t)\,,
\end{align}
where the dependence on the factorization scale $\mu$ has been suppressed for simplicity. We use the short-hand notation $\times$ to denote integral convolution over the momentum fractions $x_1,\,x_2$, as explicitly shown in Eq.~(\ref{eq:factorization}).
 The function $f_t^0$ describes the gluon splitting into a pair of massive top-quarks at the first order in $\alpha_s$ and is given by
\begin{equation}\label{eq:top-pdf-ll}
f_t^0(x,\mu) = \frac{\alpha_s}{2\pi}\log\left(\frac{\mu^2}{m_t^2}\right)\int_{x}^1 \frac{dz}{z} P_{tg}(z)\,f_g(x/z,\mu),
\end{equation}
where $P_{tg}(z)=(z^2+(1-z)^2)/2$ \cite{Altarelli:1977zs}. 
At the LO, $f_t^0(x,\mu)$ corresponds with the LL approximation of the PDF 
$f_t(x,\mu)$.\footnote{At higher orders, the corresponding functions $f_t^i$ may include factorization-scale independent terms \cite{Buza:1996wv}, which are needed when setting the boundary conditions at the heavy-quark threshold for the DGLAP evolution of the resummed PDF $f_t$.}

The organization of subtraction in Eq.~(\ref{eq:xs-acot}) is particularly suitable and intuitive for $m_{\h}\gtrsim m_t$. Essentially, we are subtracting out the LL approximation to gluon splitting found in the PDF $f_t$ in favor of the graph with explicit gluon splitting, which includes all mass effects and the full phase space. The effective PDFs $[f_t -f_t^0]$ in Eq.~(\ref{eq:xs-acot}) then approximate the contributions from higher-order collinear splittings. The numerical importance of all individual terms can be comparable, which can be understood by counting the top-quark PDF as intrinsically $\mathcal{O}(\alpha_s)$-suppressed compared to the gluon PDF. This is apparent from Eq.~(\ref{eq:top-pdf-ll}). It is also clear that LO PDFs are appropriate to use in this calculation. Once the subtraction has been performed, the effective PDF is of order $\alpha_s^2 L^2$ compared to the gluon PDF. At scales not far above the top threshold, the gluon-initiated 5-flavor LO term $gg\rightarrow t\bar t \h$ in Eq.~(\ref{eq:xs-acot}) is thus largest. Using only the 6-flavor LO term $t\bar t\to \h$ with resummed top-quark PDFs would give a very inaccurate result. 

\begin{figure}[!t]
\centering
\includegraphics[width=7.8cm]{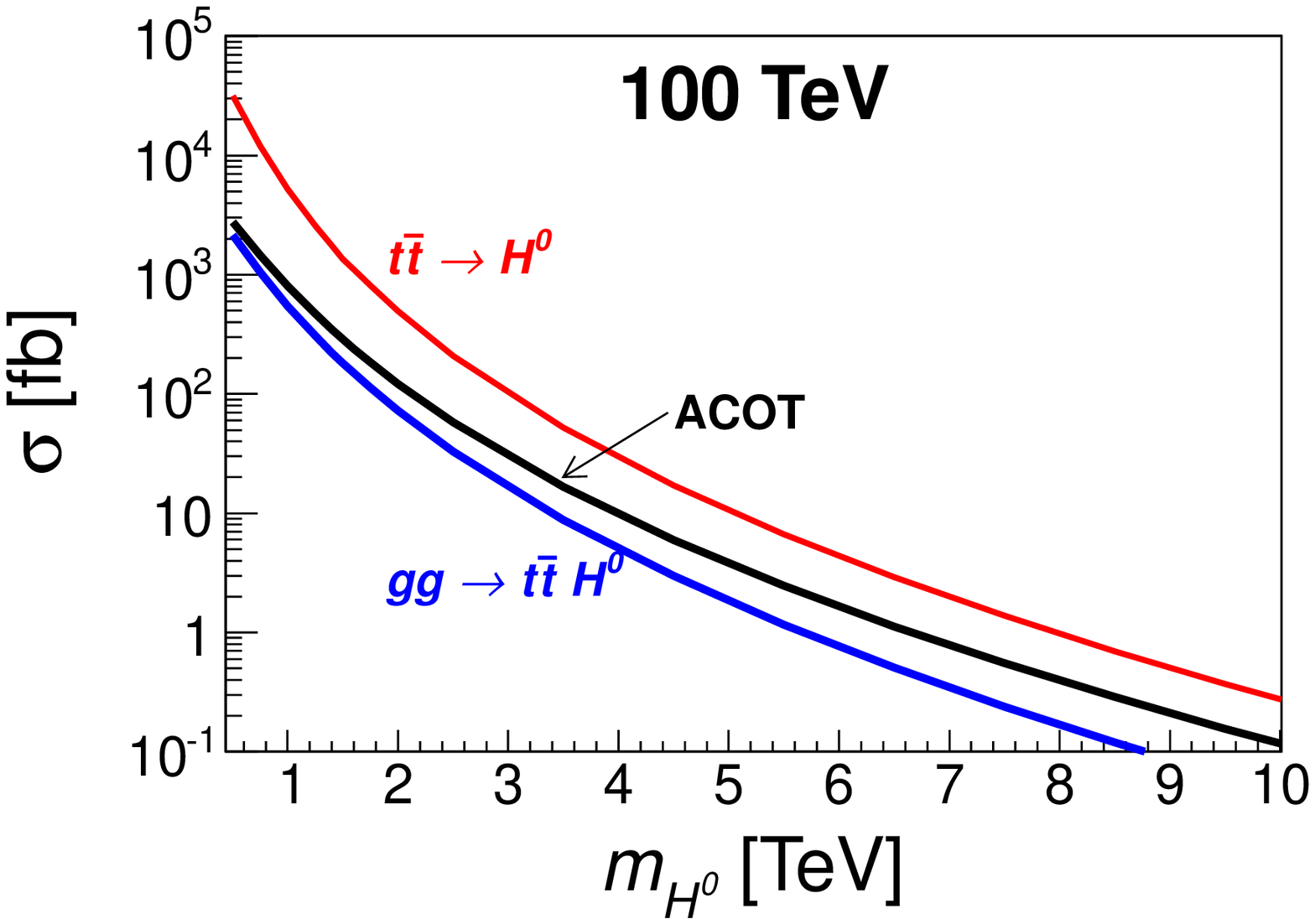}
\hspace*{0.5cm}
\includegraphics[width=7.8cm]{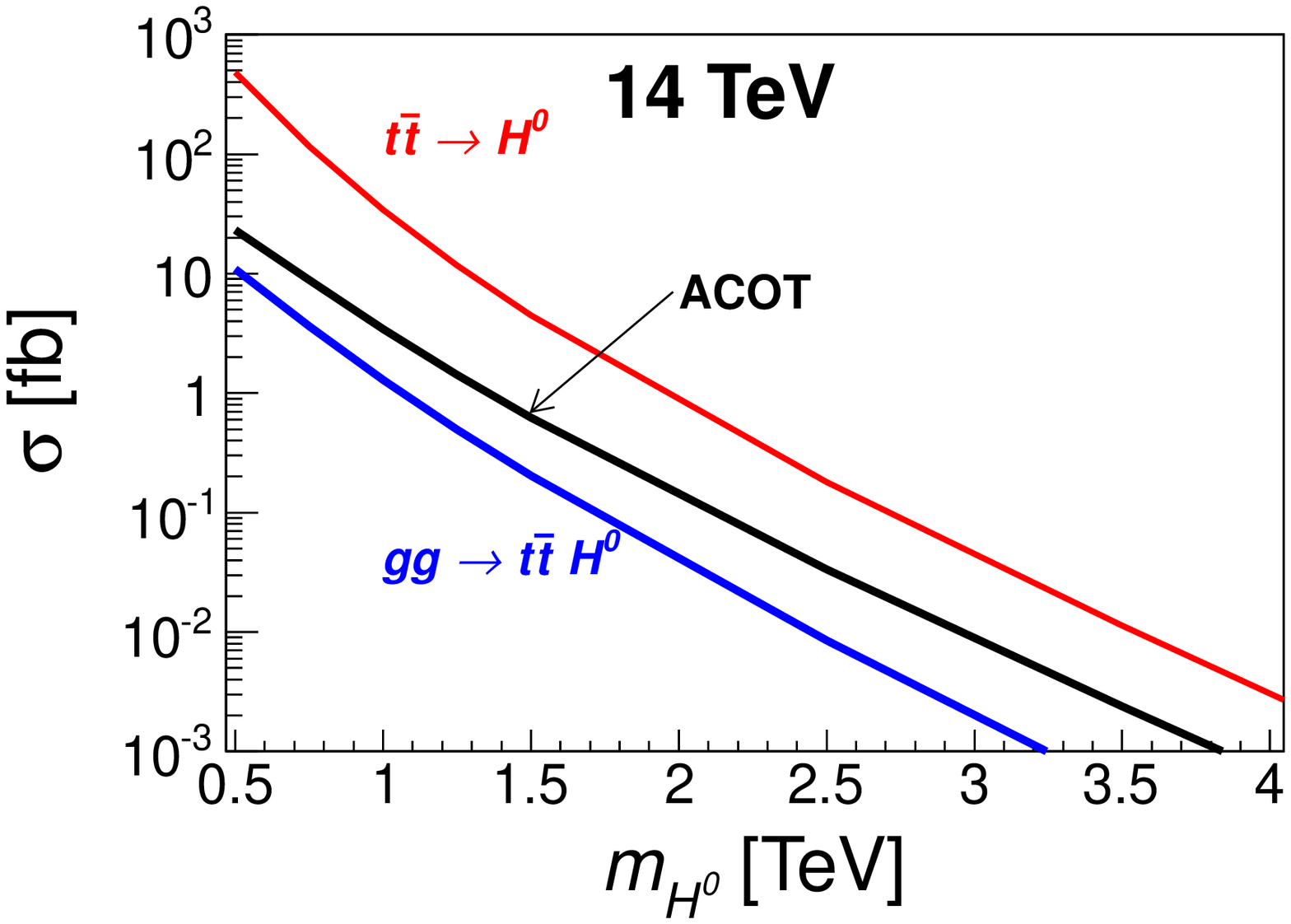}
\caption{Inclusive cross section for $\h$ production with coupling $y=1$ at $100\tev$ (left panel) and $14\tev$ (right panel) versus its mass $m_{H^0}$, in the 5-flavor scheme (bottom blue), the 6-flavor scheme (upper red), and the ACOT scheme (middle black).
}
\label{fig:schemes1}
\end{figure}

In Figure~\ref{fig:schemes1} we show the top-quark initiated inclusive cross section $\sigma_{pp\to \h}$ at tree level for proton-proton collisions at $\sqrt{S}=100\tev$ (left panel) and at $\sqrt{S}=14\tev$ (right panel). We compare the predictions in the 5-flavor scheme with massive top-quarks ($gg\to t\bar t\h$, blue), the 6-flavor scheme with massless top-quarks ($t\bar t \to \h$, red), and in the ACOT scheme defined in Eq.~(\ref{eq:xs-acot}) (black) as a function of the scalar mass $m_{\h}$. The cross section in the 6-flavor resummed scheme exceeds the 5-flavor open production by about a factor of 10 (50) at $100\tev$ ($14\tev$). As desired, the ACOT treatment interpolates smoothly between them, as seen in the figures. In particular, the excess of the 6-flavor cross section over the 5-flavor result is largely removed in ACOT at scales close to the top threshold by the LL subtraction. The remaining excess can be attributed to higher-order collinear logarithms captured by $f_t-f_t^0$. The 6-flavor prediction, on the other hand, becomes more accurate than the 5-flavor result far above the threshold at $m_{\h} \gg m_t$, where higher-order resummation plays a large role. 

The approach to the 6-flavor behavior can also be understood in terms of order-counting. Near the top threshold, processes involving $f_t$ and $f_t^0$  give individually large contributions, which will however largely cancel in the cross section. Thus it is useful to consider their sum as a correction to the gluon-initiated contributions. At high scales, $f_t^0$ grows much larger than $f_t$, and at the same time the collinear region of explicit gluon splitting comes to dominate the gluon-initiated graphs. Hence, in this limit the sum of gluon-initiated terms and their approximation by $f_t^0$ can be regarded as a higher-order correction compared to $f_t$. In this sense we recover the `standard' QCD order counting used in light-quark initiated processes.

Perhaps counterintuitively, the contributions of higher-order collinear resummation for a given mass $m_{\h}$ appear larger at $14\tev$ than at $100\tev$.\footnote{A similar effect was noted in \cite{Maltoni:2012pa} for resummation effects of bottom-quark emission in processes at the Tevatron versus the LHC.} This effect arises from the higher value of $\tau=m_{\h}^2/S$ at the LHC, which corresponds with higher average momentum fractions $x$ in the PDFs. Firstly, the difference between $f_t$ and $f_t^0$ grows at larger $x$, contributing to the observed increase in the ACOT prediction. Secondly, the 6-flavor LO cross section is relatively larger at the LHC compared to the 5-flavor cross section. We will discuss this issue further in Section~\ref{sec:scale}.

While the ACOT scheme is applicable to top-quark initiated $\h$ production for a broad range of scales above the top threshold, it can become cumbersome. In particular, keeping the top-quark mass throughout in the partonic process complicates higher-order QCD calculations.
We therefore consider if it is reasonable to neglect  the top-quark mass at certain points in the calculation.

\subsection{Massless top-quark partons: s-ACOT scheme}\label{sec:massless}
As shown in the previous section, at very high scales $\mu \approx m_H \gg m_t$, the leading term in the ACOT expansion, $t\bar t \to \h$, seems to capture the main features of the cross section for top-quark initiated processes. It is thus tempting to neglect the top-quark mass throughout, with an error formally of the order of $m_t^2/m_H^2$. The advantage of working in the massless-top limit is that higher-order calculations simplify significantly. Corrections of order $\alpha_s$ to the processes $t\bar t\rightarrow \h$ \cite{Dicus:1998hs} and $tg\rightarrow t\h$ \cite{Campbell:2002zm} have been calculated for $m_t = 0$. 
However, caution must be taken when considering the application of the massless-top limit to a broader range of scales, especially for $m_H\gtrsim m_t$ near the top threshold.

A modification of the ACOT scheme called ``simplified ACOT'' (s-ACOT in short) \cite{Collins:1998rz}, allows one to neglect the heavy-quark mass at specific places in the cross section without ruining factorization in DIS.\footnote{The general justification for the s-ACOT rule is discussed in Appendix~\ref{sec:massesapp}.} The commonly adopted prescription for treating heavy-quark mass effects in the s-ACOT scheme reads \cite{Kramer:2000hn}:
\begin{center}
{\it Heavy-quark masses may be neglected in hard processes with an initial heavy quark.}
\end{center}
Directly applying this rule to processes with two incoming heavy quarks and keeping the same terms as in the ACOT scheme in Eq.~(\ref{eq:xs-acot}), the cross section under consideration can be written as
\begin{align}\label{eq:xs-sacot}
\sigma_{pp\rightarrow \h} & = \Big\{f_t(x_1)\times\hat{\sigma}_{t\bar t\rightarrow \h}(m_t = 0)\times f_{\bar{t}}(x_2)\\\nonumber
+& \big[f_t(x_1)\times \hat{\sigma}_{tg\rightarrow t\h}(m_t = 0)\times f_g(x_2) - f_t(x_1)\times \hat{\sigma}_{t\bar t\rightarrow \h}(m_t = 0)\times \widetilde{f_{\bar{t}}^0}(x_2)\big] +(x_1 \leftrightarrow x_2)\phantom{\Big\}}\\\nonumber
+& \big[f_g(x_1)\times \hat{\sigma}_{gg\rightarrow t\bar t \h}(m_t \ne 0)\times f_g(x_2) - f_g(x_1)\times \hat{\sigma}_{gt\rightarrow t\h}(m_t = 0)\times f_t^0(x_2)\\\nonumber
-& f_{\bar t}^0(x_1)\times \hat{\sigma}_{\bar t g\rightarrow \bar t \h}(m_t = 0)\times f_g(x_2) + f_t^0(x_1)\times \hat{\sigma}_{t\bar t\rightarrow \h}(m_t = 0)\times f_{\bar{t}}^0(x_2)\big]\Big\} + (t\leftrightarrow \bar t).
\end{align}
The difference from Eq.~(\ref{eq:xs-acot}) is two-fold: Firstly, the top-quark mass is only retained in matrix elements with initial gluons, $gg\rightarrow t\bar t \h$. Secondly, the organization of collinear subtraction has been rearranged, so that the terms in brackets [\dots] are finite in the massless limit. In this form, the approximation of explicit $g\to t\bar t$ splitting by the LL PDF $f_t^0$ at high scales is apparent. The function $\widetilde{f_{\bar{t}}^0}$ in the second line of Eq.~(\ref{eq:xs-sacot}) denotes the gluon splitting into {\it massless} top-quarks, in order to cancel the collinear divergence of $\hat{\sigma}_{tg\to tH^0}(m_t=0)$ (see for instance \cite{Dicus:1998hs}). To further examine the validity of this approach, several remarks are in order:

\noindent
(1). The partonic cross section $\hat{\sigma}_{gg\to t\bar t \h}$ has single and double collinear divergences in the massless-top limit. Thus the divergences in the third and fourth lines of Eq.~(\ref{eq:xs-sacot}) cannot be cancelled, if the top mass is retained in $\hat{\sigma}_{gg\to t\bar t \h}$, but not in $\hat{\sigma}_{tg\to t\h}$ and $\hat{\sigma}_{\bar tg\to \bar t\h}$.
We will therefore retain the top-quark mass in all three contributions, which seemingly conflicts with the s-ACOT rule. According to the arguments of the previous section, these terms can be neglected as a higher-order correction to $t\bar t\to \h$ at very high scales, but should be included in general.

\noindent
(2). Consider the process $tg\rightarrow t\h$ together with its subtraction term, as in the second line of Eq.~(\ref{eq:xs-sacot}). Using the short-hand notation $\hat{\overline{\sigma}}_{tg\to t \h}$ to denote the remainder after subtraction, it can be expressed as \cite{Dicus:1998hs}
\begin{align}\label{eq:massless}
(f_t\times \hat{\overline{\sigma}}_{tg\rightarrow t\h}\times f_g)_{m_t=0} & = \frac{\alpha_s}{24}\frac{y^2}{S}\int_{\tau}^1\frac{\text{d}x}{x}\,f_t(x,\mu)\int_{\tau/x}^1\frac{\text{d}z}{z}\,f_g\Big(\frac{\tau}{zx},\mu\Big)\\\nonumber
&\quad \times \Big[P_{tg}(z)\log\left(\frac{m_{\h}^2}{\mu^2}\frac{(1-z)^2}{z}\right) + \frac{1}{4}(1-z)(7z-3)\Big]\,,
\end{align}
where $y$ is the Yukawa coupling in Eq.~(\ref{eq:coupling}). 
 The collinear divergence cancels in the massless limit, and the logarithm $\log(m_{\h}^2/\mu^2)$ is a remnant of this cancellation. The mass ratio suggests $\mu\approx m_{\h}$ as a natural choice of factorization scale to reduce scale effects, as adopted in Eq.~(\ref{eq:mu}).

\noindent
(3). We can identify $z\equiv m_{\h}^2/s$ in Eq.~(\ref{eq:massless}) with the momentum fraction carried by an incoming massless top-quark from gluon splitting in the collinear limit. The additional factor of $\log[(1-z)^2/z]$ can be large and negative for $z\to 1$, where $\h$ is being produced close to threshold and the outgoing top-quark carries a small fraction $(1-z)$ of the gluon momentum. In reality, the finite top-quark mass sets an upper bound $z_{\rm max} =m_{\h}^2/(m_{\h}+m_t)^2$ in the underlying physical process $tg\to t\h$. The massless-top limit, however, leads to integration over an unphysical region,
\begin{equation}
z_{\rm max} \approx 1- {2 m_t\over m_{\h}} \lesim \tilde{z} \le 1.
\label{eq:zunphy}
\end{equation}
The difference between the massless-top limit and the (massive) ACOT approach for this process $tg\rightarrow t\h$ can be expressed by 
\begin{align}\label{eq:massless2}
& (f_t\times \hat{\overline{\sigma}}_{tg\rightarrow t\h}\times f_g)_{m_t=0} - (f_t \times \hat{\overline{\sigma}}_{tg\rightarrow t\h}\times f_g)_{m_t} =\\\nonumber
& \quad \frac{\alpha_s}{24}\frac{y^2}{S}\int_{z_{\rm max}}^{1} \frac{\text{d}\tilde{z}}{\tilde{z}}\int_{\tau/\tilde{z}}^{1}\frac{\text{d}x}{x} f_t(x) f_g\left(\frac{\tau}{\tilde{z}x}\right) \Big[P_{tg}(\tilde{z})\log\left(\frac{m_{H^0}^2}{m_t^2}\frac{(1-\tilde{z})^2}{\tilde{z}}\right) +\mathcal{O}(1-\tilde{z})\Big].
\end{align}
This unphysical error should be compared to the physical cross section 
\begin{align}
\sigma_{pp \to H^0} \approx f_t \times (\hat \sigma_{t \bar t \to H^0} \times [f_{\bar t}-f_{\bar t}^0] + \hat \sigma_{tg \to tH^0} \times f_g). 
\label{eq:oneside}
\end{align}
Near the top threshold, we find that the error Eq.~(\ref{eq:massless2}) is negative and larger than the physical cross section.\footnote{This large negative correction has been observed in \cite{Dicus:1998hs}, but no further investigation was offered.} However,
as $m_t/m_{H^0} \to 0$ the error dwindles in comparison to $(f_t \times \hat \sigma_{tg \to t\h} \times f_g)$. This is due to 
$z_{max} \to 1$ in Eq.~(\ref{eq:massless2}) eliminating the unphysical region of integration and the factor $m_{H^0}^2/m_t^2$ offsetting the negative contribution in the logarithm. Furthermore, the growing effective PDF $[f_t-f_t^0]$ mitigates the impact of $tg\to tH^0$ on the total cross section.

\begin{figure}[!t]
\centering
\includegraphics[width=9cm]{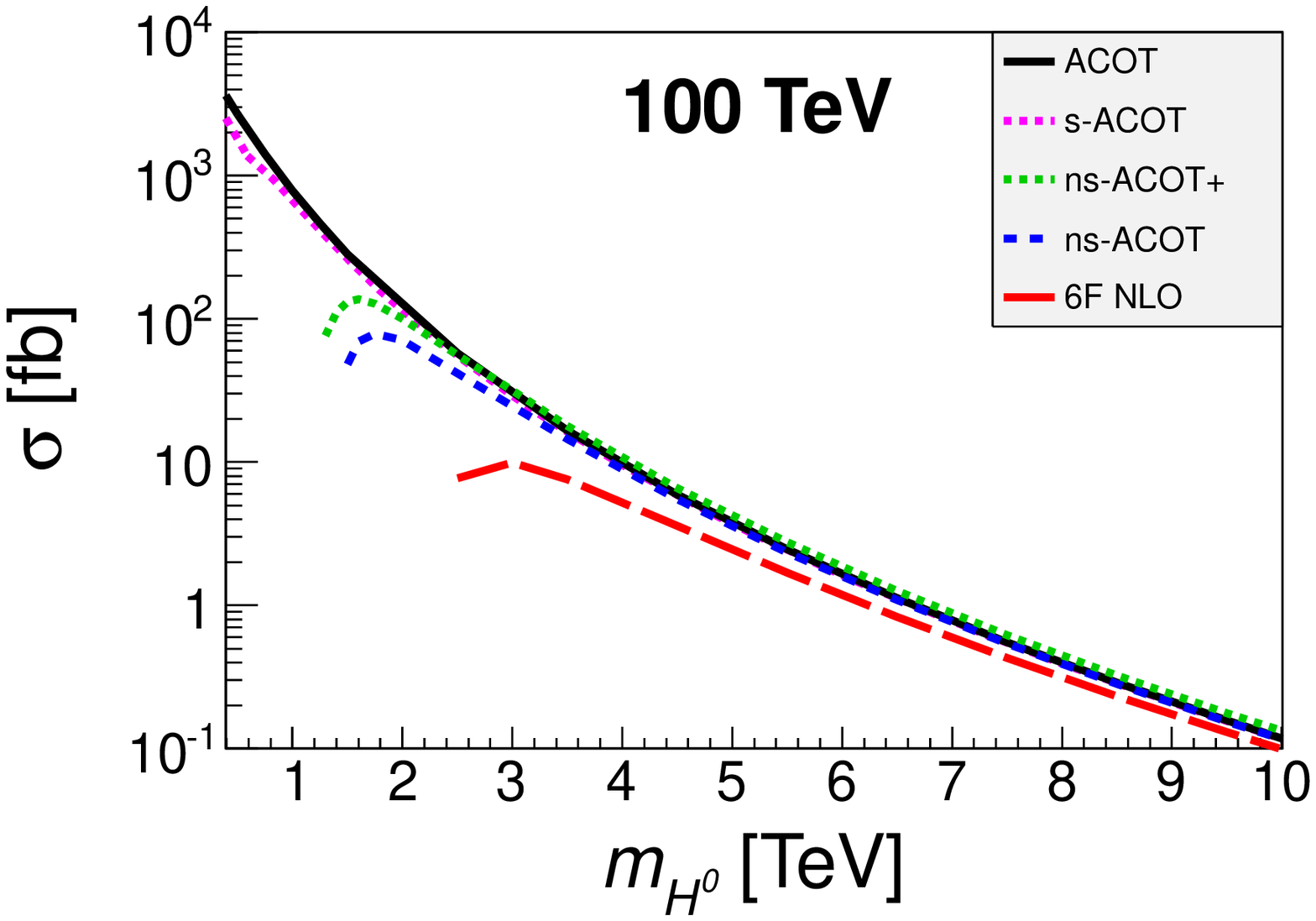}
\caption{
Inclusive cross section for $\h$ production with coupling $y=1$ at $100\tev$ versus its mass $m_{\h}$, in the ACOT scheme (top black), the s-ACOT (magenta dotted curve, Eq.~(\ref{eq:xs-sacot-adjusted})), the naive s-ACOT result with NLO corrections labelled as ns-ACOT+ (green dotted curve), 
the naive s-ACOT result at LO (blue dotted curve, labelled ns-ACOT), and the massless 6-flavor calculation at NLO (bottom red).
}
\label{fig:zerom}
\end{figure}
An improved treatment of processes with two initial heavy quarks in the s-ACOT scheme has been suggested \cite{Collins&Yuan}. In this approach, the divergences in the third line of Eq.~(\ref{eq:xs-sacot}) are canceled by including additional subtraction terms, while taking the massless limit of all heavy-quark initiated graphs. Whenever neglecting the heavy-quark mass in the subtraction terms induces a collinear divergence, this divergence needs to be subtracted from the kernel. The cross section for top-initiated scalar production can then be expressed as
\begin{align}\label{eq:xs-sacot-adjusted}
\sigma_{pp\rightarrow \h} & = \Big\{f_t(x_1)\times\hat{\sigma}_{t\bar t\rightarrow \h}(m_t = 0)\times f_{\bar{t}}(x_2)\\\nonumber
+& \big[f_t(x_1)\times \hat{\sigma}_{tg\rightarrow t\h}(m_t = 0)\times f_g(x_2) - f_t(x_1)\times \hat{\sigma}_{t\bar t\rightarrow \h}(m_t = 0)\times \widetilde{f_{\bar{t}}^0}(x_2)\big] +(x_1 \leftrightarrow x_2)\phantom{\Big\}}\\\nonumber
+& f_g(x_1)\times \hat{\sigma}_{gg\rightarrow t\bar t \h}(m_t \ne 0)\times f_g(x_2)\\\nonumber
&\quad - \big[f_g(x_1)\times \hat{\sigma}_{gt\rightarrow t\h}(m_t = 0) - \widetilde{f_{\bar{t}}^0}(x_1)\times \hat{\sigma}_{\bar t t\rightarrow \h}(m_t = 0) \big]\times f_t^0(x_2) + (x_1\leftrightarrow x_2)\\\nonumber
&\quad - f_{\bar t}^0(x_1)\times \hat{\sigma}_{\bar t t\rightarrow \h}(m_t = 0)\times f_{t}^0(x_2)\Big\} + (t\leftrightarrow \bar t).
\end{align}
When masses are restored to all terms this expression will reduce to the ACOT formula.
In this form, the LO cross section agrees with the ACOT prediction up to corrections of the order $\alpha_s^3 L^2\cdot (m_t^2/m_{\h}^2)$, which are formally suppressed by the strong coupling constant. In particular, the impact of the unphysical region $z_{\rm max} < \tilde{z} < 1$ of soft-top emission in the process $tg\to t\h$ is therefore less significant than in the naive version of s-ACOT illustrated earlier. However, for scalar masses not far above the top threshold, the cross section still suffers from neglecting the top-quark mass in the region of soft-top emission.

To see the behavior of the different schemes more explicitly, we compare predictions of the total cross section $\sigma_{pp\to \h}$ in four schemes in Figure~\ref{fig:zerom}. The (black) solid curve reproduces the ACOT result of Eq.~(\ref{eq:xs-acot}) presented earlier. The (red) dashed line represents the massless 6-flavor scheme at next-to-leading order (NLO), corresponding to the first two lines in Eq.~(\ref{eq:xs-sacot}), along with virtual and real corrections to $f_t\times \hat{\sigma}_{t\bar t \rightarrow \h}\times f_{\bar t}$. As expected the massless 6-flavor scheme agrees with the ACOT prediction for large masses of the heavy particle $\h$, but deviates significantly for $m_{\h} \lesssim 5\tev$ and plummets around $3\tev$, indicating a clearly unphysical result. The cross section in the naive s-ACOT scheme of Eq.~(\ref{eq:xs-sacot}) (with $m_t$ retained in lines 3 and 4) is shown by the (blue) short-dashed curve. By keeping $gg\rightarrow t\bar t \h$ and its subtraction terms, the region of validity extends to $m_{\h} \gtrsim 3\tev$. Nevertheless, the error from neglecting $m_t$ in the process $tg\to t\h$ is significant below this scale and becomes pathological near $m_{\h} \approx 2\tev$. To further consolidate our observation, we also show the naive s-ACOT result with NLO corrections to $f_t\times \hat{\sigma}_{t\bar t \to \h}\times f_{\bar t}$ included (labelled as ns-ACOT+), indicated by the (green) dotted curve. These  corrections are positive and can slightly improve the agreement with the ACOT prediction, but ultimately they do not cure the unphysical behavior. The s-ACOT prediction from Eq.~(\ref{eq:xs-sacot-adjusted}) is shown by the (magenta) dotted curve. This s-ACOT prediction agrees with the ACOT result apart from the region close to the top threshold. In this region the impact of the integration region $z_{\rm max} < \tilde{z} < 1$ leads to a negative (unphysical) correction to the 5-flavor LO contribution.

\subsection{Modified mass treatment: m-ACOT scheme}\label{sec:macot}
In light of the preceding discussion, the s-ACOT prescription presents some ambiguities when applied to the full range of energy scales. The key difficulty with s-ACOT is that taking the massless limit $m_t\to 0$ in a hard-scattering graph with one initial gluon will generically allow large contributions from unphysically small virtuality in $t$-channel propagators. {This effect is moderated in the s-ACOT scheme with additional terms described above, but still causes a significant deviation from the ACOT prediction. We therefore propose a modification of ACOT, which we call m-ACOT. It preserves some benefits of simplified higher-order calculations without the problematic behavior in the region of collinear and soft top-quark emission. The m-ACOT prescription for processes with heavy-quark fusion reads:
\begin{center}
{\it The heavy-quark mass may be neglected only\\
in partonic processes with two incoming heavy quarks.}
\end{center}
Masses must be kept in all processes with at least one initial gluon. The total cross section for the $t\bar t$-initiated production of a neutral boson $H^0$ takes the form 
\begin{align}\label{eq:xs-sacot-neutral}
\sigma_{pp\rightarrow H^0} & = \Big\{[f_t-f_t^0](x_1)\times \hat{\sigma}_{t\bar t\rightarrow H^0}(m_t=0)\times [f_{\bar t}-f_{\bar{t}}^0](x_2) \\\nonumber
&\ \ \, + [f_t-f_t^0](x_1)\times \hat{\sigma}_{tg\rightarrow tH^0}(m_t \ne 0)\times f_g(x_2) + (x_1\leftrightarrow x_2) \\ \nonumber
&\ \ \, + \qquad\ \ f_g(x_1)\times \hat{\sigma}_{gg\rightarrow t\bar t H^0}(m_t \ne 0)\times f_g(x_2)\Big\}  + (t\leftrightarrow \bar t).
\end{align}
The modification with respect to the ACOT formalism in Eq.~(\ref{eq:xs-acot}) is simply setting $m_t=0$ in the $t\bar t$-initiated contributions. This formalism also applies to the analogous $b\bar b$-initiated processes at scales $m_{\h} > m_b$.

For the $t\bar b$-initiated production of a charged boson $H^+$, the bottom-quark mass can be set to zero throughout. The cross section in m-ACOT thus reads 
\begin{align}\label{eq:xs-sacot-charged}
\sigma_{pp\rightarrow H^+} & = \Big\{[f_t-f_t^0](x_1)\times \hat{\sigma}_{t\bar b\rightarrow H^+}(m_t=0,m_b=0)\times [f_{\bar b}-\widetilde{f_{\bar{b}}^0}](x_2) \\\nonumber
&\ \ \, + [f_{\bar b}-\widetilde{f_{\bar b}^0}](x_1)\times \hat{\sigma}_{\bar{b}g\rightarrow \bar{t}H^+}(m_t \ne 0,m_b=0)\times f_g(x_2) + (x_1\leftrightarrow x_2)\\\nonumber
&\ \ \, + [f_t-f_t^0](x_1)\times \hat{\sigma}_{tg\rightarrow bH^+}(m_t=0,m_b=0)\times f_g(x_2) + (x_1\leftrightarrow x_2)\phantom{\Big\}}\\\nonumber
&\ \ \, + \qquad\ \ f_g(x_1)\times \hat{\sigma}_{gg\rightarrow b\bar t H^+}(m_t \ne 0,m_b=0)\times f_g(x_2) 
\Big\}\, + (t\leftrightarrow \bar b)
,
\end{align}
and similarly for $H^-$ with $t\leftrightarrow \bar t$ and $b\leftrightarrow \bar b$. In the $\bar b g$-initiated contributions in Eq.~(\ref{eq:xs-sacot-charged}), the virtuality of the $t$-channel propagator is regularized by the top-quark mass. We stress that the treatment of heavy-quark masses in Eq.~(\ref{eq:xs-sacot-neutral}) and Eq.~(\ref{eq:xs-sacot-charged}) can be generalized to higher orders of $\alpha_s$ in QCD. \footnote{As a caveat, we note that to our knowledge a proof of the validity of the ACOT scheme for hadron-hadron collisions to all orders has not been presented in the literature.} In particular, new channels with initial light quarks opening up at higher orders do not affect our mass treatment of heavy quarks. Due to the large hierarchy between the top-quark and light-quark masses, light quarks can be treated as massless at energies above the top production threshold to a very good approximation.

In our example process $pp\to\h$, neglecting the top-mass in $\hat{\sigma}_{t\bar t\to \h}$ at the LO is certainly a small modification from the calculational perspective. The numerical difference between the LO cross section $\sigma_{pp\to H^0}$ in the proposed m-ACOT and the original ACOT scheme as in Figure~\ref{fig:schemes1} is less than one percent for $m_{\h} \gesim m_t$. As desired, factorization is preserved in m-ACOT without introducing corrections of order $m_t^2/m_{H^0}^2$. Beyond LO, the m-ACOT simplification implies that the top-quark mass can be neglected in QCD corrections to $t\bar t \rightarrow \h$ from virtual and real final-state gluon radiation to all orders in $\alpha_s$. This is useful especially in the regime $m_{H^0}\gg m_t$, because higher-order corrections to $tg\rightarrow t\h$ and $gg\rightarrow t\bar t \h$ are numerically subleading. We will discuss this issue in more detail in Section~\ref{sec:dis}.


\section{Neutral heavy-scalar production in m-ACOT}\label{sec:vlhc}
In the numerical analysis of our process $pp\to\h$, we will make use of the m-ACOT scheme at LO advocated in Eq.~(\ref{eq:xs-sacot-neutral}) throughout the remainder of this work.
 Our results have been calculated with two independent private computer codes including MadGraph components \cite{madgraph}. The phase-space integration has been performed numerically using the integration routine Vegas from the Cuba library \cite{Hahn:2004fe}. We have generated PDFs above the top-mass scale as described in Section~\ref{sec:proc}.

\begin{figure}[!t]
\centering
\mbox{\subfigure{
\includegraphics[angle=0,scale=0.4]{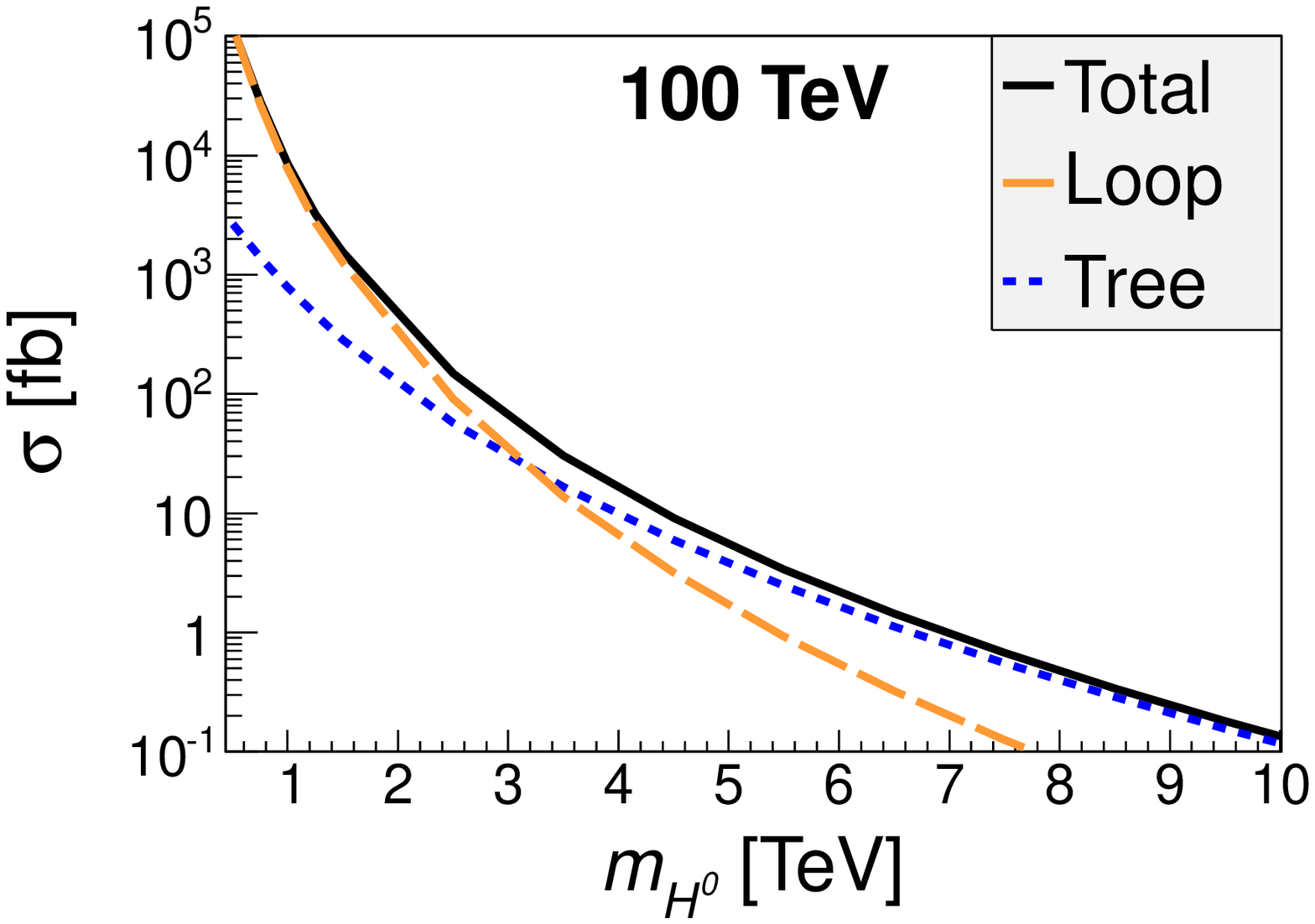}} \subfigure{
\includegraphics[angle=0,scale=0.4]{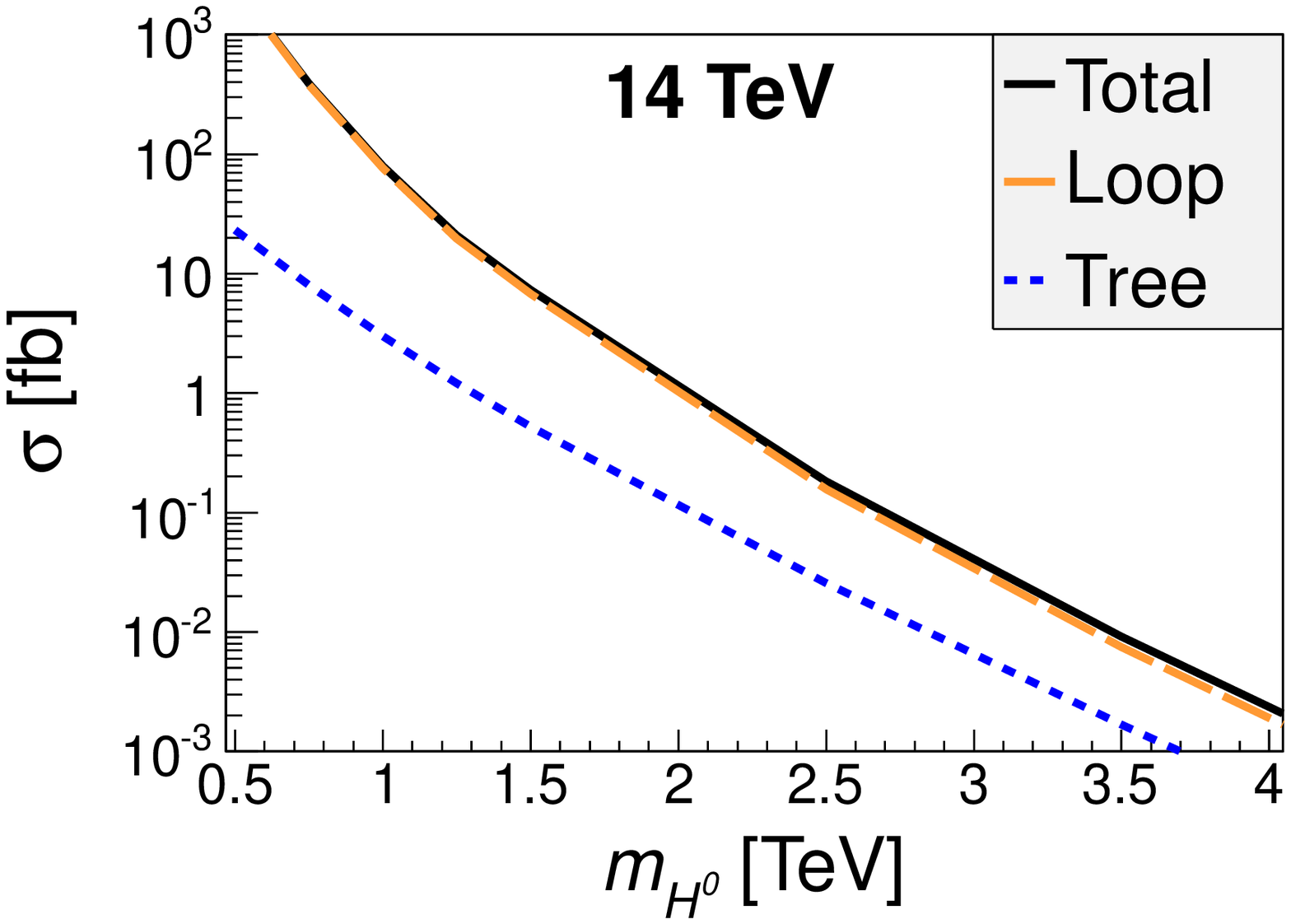}}}
\caption[]{Inclusive LO cross section for $H^0$ production at 100 TeV (left panel) and 14 TeV (right panel) versus its mass. Shown are the tree-level contributions in the m-ACOT scheme (blue short-dashed curves), $gg$ fusion via a top loop (orange long-dashed curves), and the sum of both contributions (black solid curves).}
\label{fig:lhc-fcc}
\end{figure}

There is one potentially important term we have neglected to discuss thus far: the production of a heavy scalar from gluon-gluon ($gg$) fusion via a triangular top-quark loop
\begin{align}
\label{eq:xs-gg}
gg\rightarrow H^0.
\end{align}
The contribution of gluon-gluon fusion to the total cross section is of order $\alpha_s^2$ and thus of the same order as the  tree-level contributions, counting $f_t$ as $\alpha_s$-suppressed. In Figure~\ref{fig:lhc-fcc} we show the total cross section for $t\bar t$-initiated scalar production in the m-ACOT scheme (Eq.~(\ref{eq:xs-sacot-neutral})) and identify the contribution from $gg$ fusion in Eq.~(\ref{eq:xs-gg}). At the $14$-TeV LHC (right panel), $gg$ fusion dominates the cross section. The tree-level contributions give a correction of up to $\mathcal{O}(10\%)$ for $m_{\h} < 4\tev$. At a $100\tev$-collider (left panel), however, the $gg$ fusion cross section falls faster over $m_{\h}$, and the tree-level processes take over above $m_{\h} \sim 3\tev$.

When considering scalar production in association with one or more tops, the loop-induced process $gg$ fusion does not contribute at LO, and the cross section is dominated by the tree-level processes $tg\to t\h$ and $gg\to t\bar t \h$. For the inclusive production of a heavy vector boson as listed in Table~\ref{tab:sqme}, the contribution from $gg$ fusion will be vanishingly small \cite{Dicus:1988cx}.

\subsection{Dependence on the factorization scale}\label{sec:scale}
When treating the top-quark as a parton, the LO 6-flavor process $t\bar t\to \h$ 
is strongly dependent on the factorization scale. Choosing the factorization scale $\mu=m_{\h}$ may be motivated by cancelling the strong logarithmic dependence $\log(m_{\h}^2/\mu^2)$ in Eq.~(\ref{eq:massless}). However, it has been argued that in the analogous process of Higgs boson production from $b\bar b$ fusion, $b\bar b\to \h$, the appropriate factorization scale is significantly lower than the Higgs mass \cite{Maltoni:2003pn}. Here we examine this issue for top-quark initiated processes. 

\begin{figure}[!t]
\centering
\includegraphics[width=7.5cm]{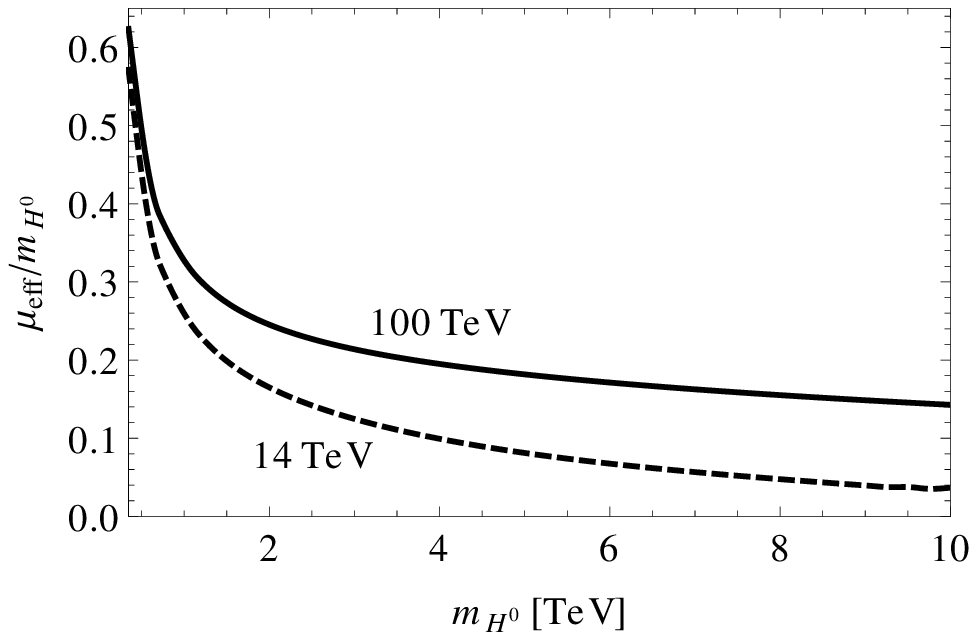}
\hspace*{0.5cm}
\includegraphics[width=7.5cm]{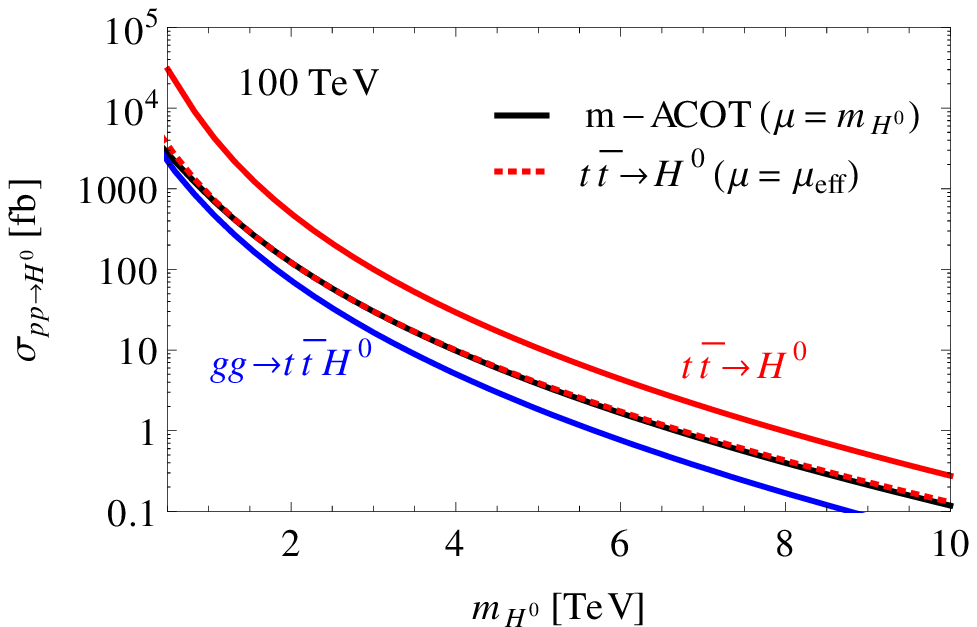}
\caption{Left: Effective factorization scale $\mu_{\rm eff}$ from Eq.~(\ref{eq:dyn-scale}) divided by $m_{\h}$ for $\sqrt{S}=100\tev$ (solid) and $\sqrt{S}=14\tev$ (dashed).
Right: The LO total cross section $\sigma_{pp\to \h}$ at $\sqrt{S}=100\tev$ as a function of $m_{\h}$ in the massless 6-flavor scheme using the effective factorization scale $\mu = \mu_{\rm eff}$ (dotted red), compared with the default scale choice $\mu = m_{\h}$ in the massless 6-flavor scheme (solid red), the 5-flavor scheme (blue) and m-ACOT (black).
}
\label{fig:xs-dyn}
\end{figure}

The essence of the use of a top-quark PDF resides in the gluon splitting into a collinear $t \bar t$ pair. Following the arguments in \cite{Maltoni:2012pa}, one can choose an effective scale in $g\rightarrow t\bar t$ splitting to make the LL approximation match the full matrix element calculation with explicit gluon splitting. In the m-ACOT scheme, applying the LL approximation to one of the initial top-quarks in the leading process $t \bar t\rightarrow \h$, we encounter the typical logarithmic dependence on the factorization scale $\mu$, 
\begin{align}\label{eq:ll}
f_t\times \hat{\sigma}_{t \bar t \rightarrow \h}\times f_{\bar t}^0 = \frac{\alpha_s}{24}\frac{y^2}{S}\log\left(\frac{\mu^2}{m_t^2}\right)\int_{\tau}^1\frac{\text{d}x}{x}\,f_{t}(x,\mu)\int_{\tau/x}^1\frac{\text{d}z}{z}\,P_{tg}(z)\,f_g\Big(\frac{\tau}{zx},\mu\Big),
\end{align}
This expression should correspond to the collinear region of the process $tg\rightarrow t\h$, which is given by the logarithmic term in Eq.~(\ref{eq:massless}) with $\mu^2\rightarrow m_{t}^2$. We can thus define an ``effective factorization scale'' $\mu_{\rm eff}$ by matching the LL approximation onto the full result in the collinear limit,
\begin{align}\label{eq:dyn-scale}
\log\left(\frac{\mu_{\rm eff}^2}{m_t^2}\right) = \frac{\int \frac{\text{d}x}{x}\,f_t(x,m_{\h})\int \frac{\text{d}z}{z}P_{tg}(z)\log(\frac{m_{\h}^2}{m_t^2}\frac{(1-z)^2}{z})f_g(\frac{\tau}{zx},m_{\h})}{\int \frac{\text{d}x}{x}\,f_t(x,m_{\h})\int \frac{\text{d}z}{z}P_{tg}(z)f_g(\frac{\tau}{zx},m_{\h})}\,.
\end{align}
Here we have used $\mu=m_{\h}$ as an input scale for the PDFs. The so-obtained effective factorization scale $\mu_{\rm eff}$ is displayed in the left panel of Figure~\ref{fig:xs-dyn}. We show the ratio 
$\mu_{\rm eff}/m_{\h}$ as a function of $m_{\h}$ for CM energies of $100\tev$ (solid) and $14\tev$ (dashed).
The effective scale $\mu_{\rm eff}$ is indeed significantly reduced with respect to the scalar mass. Especially for large $m_{\h}\gtrsim 1.5\tev$, $\mu_{\rm eff}$ is reduced to below $30\%$ ($\sqrt{S}=100\tev$) and $20\%$ ($\sqrt{S}=14\tev$) of the scalar mass.  This can be seen from Eq.~(\ref{eq:dyn-scale}). As the momentum fraction $z=m_{\h}^2/s$ becomes large, the factor controlling the collinear logarithm is $m_{\h}^2 (1-z)^2$, significantly smaller than the naive expectation $m_{\h}^2$. For a fixed CM energy, the scale reduction is thus stronger for a heavy boson. Conversely, for a fixed mass $m_{\h}$, $z$ is on average larger at $\sqrt{S}=14\tev$ than at $100\tev$, which explains why the reduction is more pronounced at the LHC. 

With the hope for improvement using the effective scale, we show the result for the LO total cross section $\sigma_{pp\rightarrow \h}$ at $\sqrt{S}=100\tev$ in Figure~\ref{fig:xs-dyn}, right panel. It is impressive to see that, with the choice $\mu_{\rm eff}$, the simple calculation for $t \bar t\rightarrow \h$ in the 6-flavor scheme (red dotted curve) reaches an excellent agreement with the full m-ACOT prediction (black solid curve) using $\mu = m_{\h}$. For comparison, predictions for $\mu=m_{\h}$ in the 6-flavor scheme (red) and the 5-flavor scheme (blue) are also shown.

\begin{figure}[!]
\centering
\includegraphics[width=7.5cm]{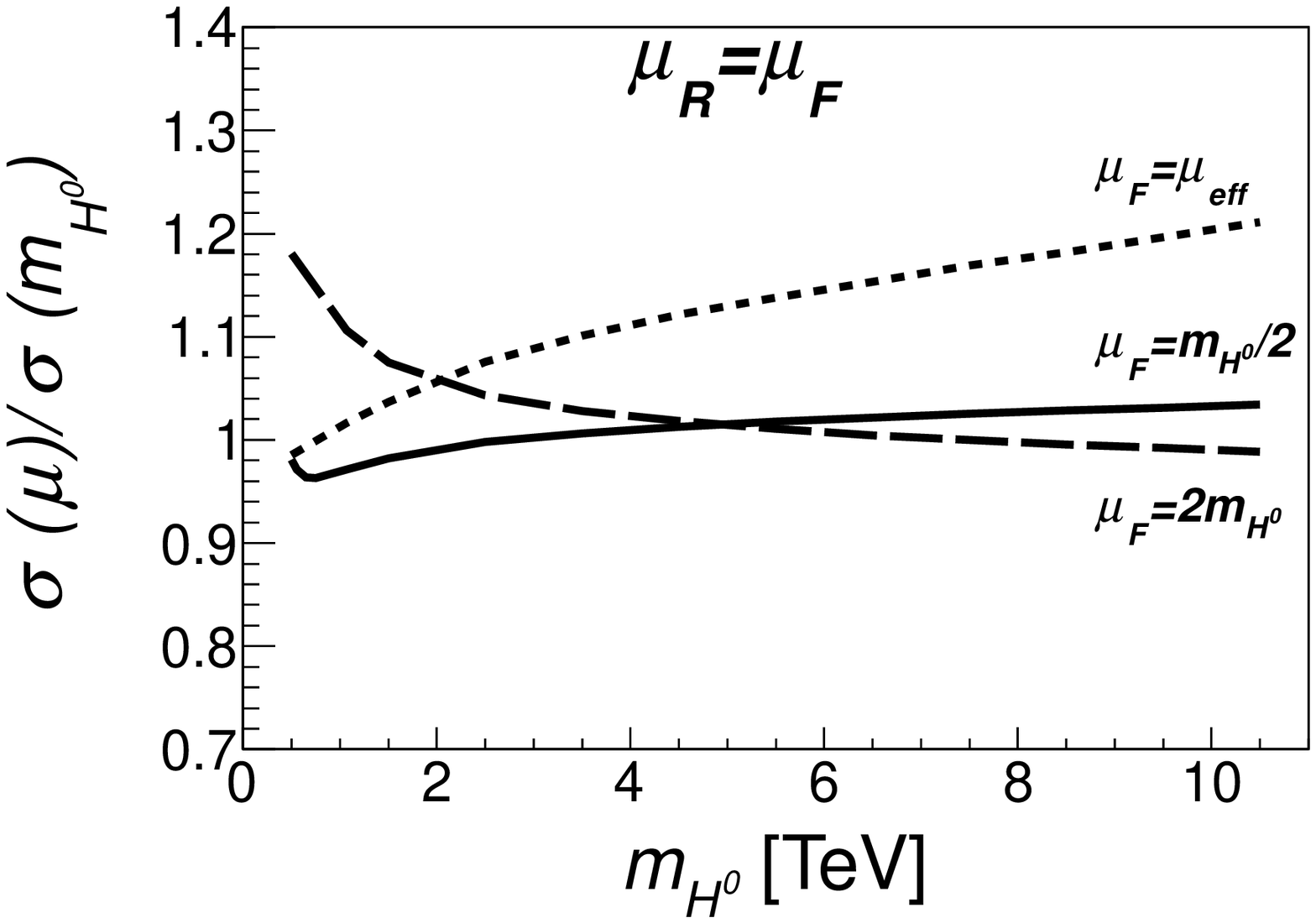}
\hspace*{0.5cm}
\includegraphics[width=7.5cm]{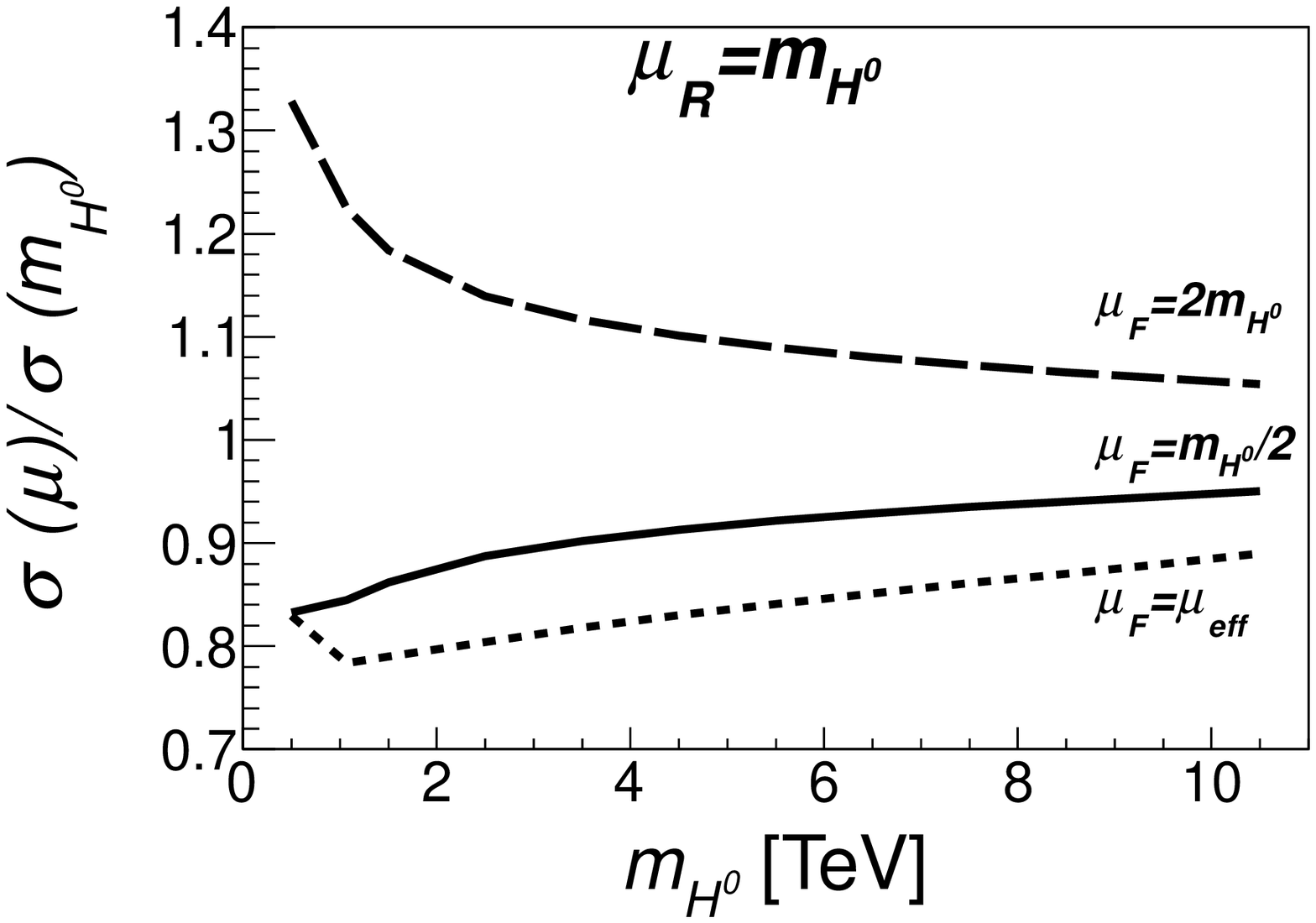}
\caption{The LO total cross section $\sigma_{pp\rightarrow \h}(\mu)$ in m-ACOT for different factorization scale choices $\mu_F=m_{\h}/2$ (solid), $2m_{\h}$ (dashed), $\mu_{\rm eff}$ (dotted), normalized to the cross section with $\mu_F = m_{\h}$. The factorization scale is set to $\mu_R=\mu_F$ (left panel) and $\mu_R=m_{\h}$ (right panel).
}
\label{fig:scale-variation}
\end{figure}

In general, the cross section in the m-ACOT framework is less scale-dependent than the 6- or 5-flavor LO calculations. In Figure~\ref{fig:scale-variation}, we show the ratio between the cross section with several scale choices $\mu = m_{\h}/2$ (solid), $2m_{\h}$ (dashed), $\mu_{\rm eff}$ (dotted) and the cross section with the default scale $\mu=m_{\h}$ versus the scalar mass $m_{\h}$ at 100 TeV.
 In the left panel, we set the renormalization scale to be the same as the factorization scale $\mu_R= \mu_F=\mu$. For the range of $m_{\h}$ under consideration, there is at most a difference of about $20\%$ in our predictions. The rather modest scale dependence is due to two factors. Firstly, the collinear subtraction in m-ACOT greatly reduces the large scale dependence of the LO 6-flavor prediction, especially for $m_{\h}$ near the top threshold. Secondly, stability is helped by using the same scale for factorization and renormalization: 
As $\mu_R$ increases, $\alpha_s(\mu_R)$ decreases, reducing the contributions with inital gluons, $tg \to t\h$ and $gg \to t \bar t \h$. At the same time, as $\mu_F$ increases, $f_t(x,\mu_F)$ increases, 
enhancing the top-quark initiated contributions relative to the gluons. These two effects offset each other in the total cross section, contributing to a more stable prediction. To separate these effects, we show the same ratios of cross sections with a separate renormalization scale $\mu_R=m_{\h}$ in the right panel. The remnant factorization-scale dependence is larger than that in the left panel as expected. In the right panel, increasing the factorization scale increases the cross section due to resummation effects and a larger gluon PDF. In the left panel, this is counteracted by the decreasing size of $\alpha_s$, especially at high $m_{\h}$. The opposite effect applies to a reduced scale. Note that the dependence of the cross section on $\mu_F$ is largest close to the top threshold, where the evolution of $f_t(x,\mu_F)$ is strongest.

\subsection{Kinematical Distributions}\label{sec:distributions}
Going beyond the LO process $t\bar t \to \h$, the $2\to 2$ process $t g \to t \h$ and the $2\to 3$ process
$gg \to t \bar t \h$ lead to interesting kinematical features. We consider three physical observables: the invariant mass of the heavy scalar and the top-quark ($M_{\h t}$), as well as the transverse momenta of the top-quark ($p_T^t$) and the heavy scalar ($p_T^{\h}$).
\begin{figure}[t]
\centering
\begin{tabular}{c}
\raisebox{4.5cm}{(a)}\includegraphics[width=7.7cm]{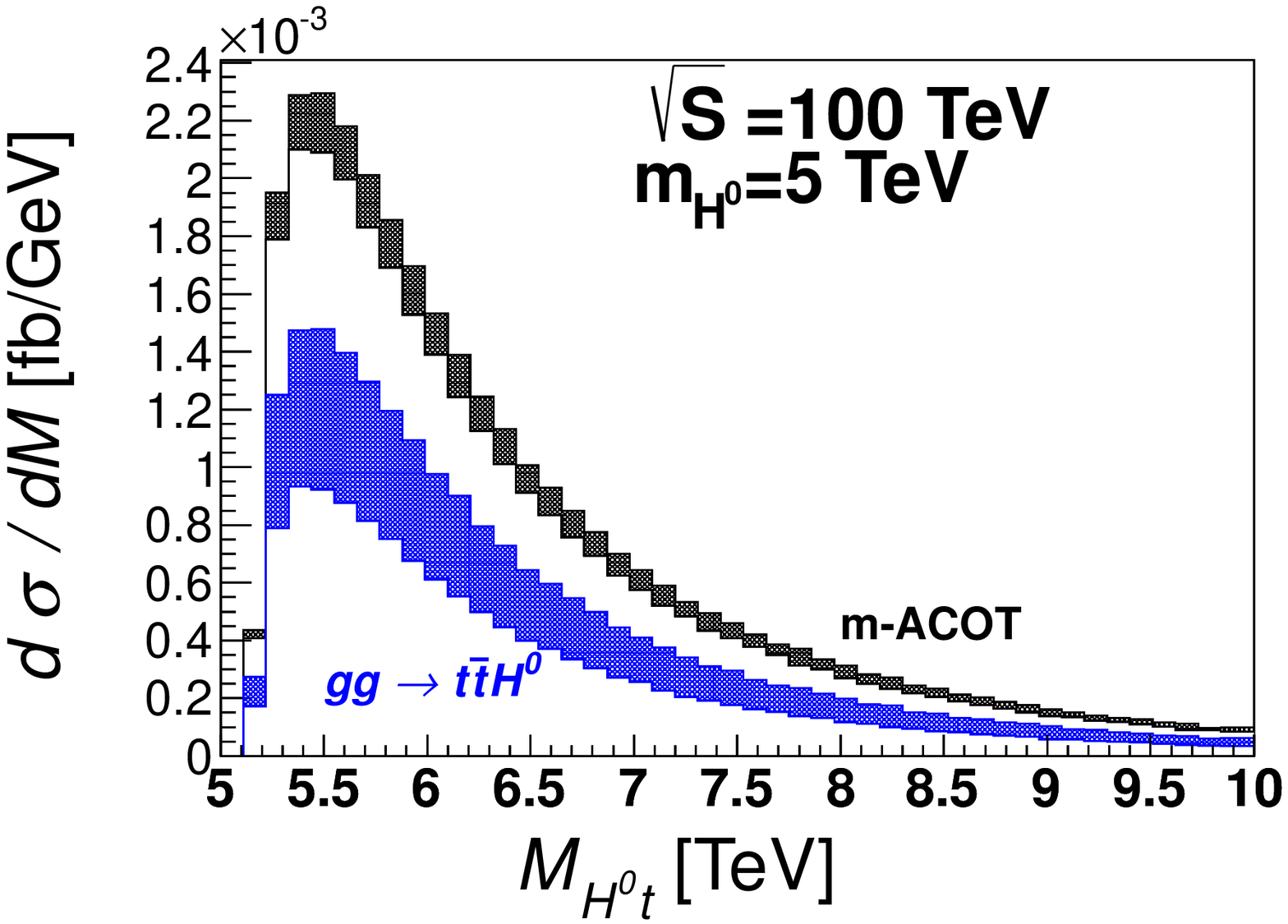}
\end{tabular}
\begin{tabular}{cc}
\raisebox{4.5cm}{(b)}\includegraphics[width=7.7cm]{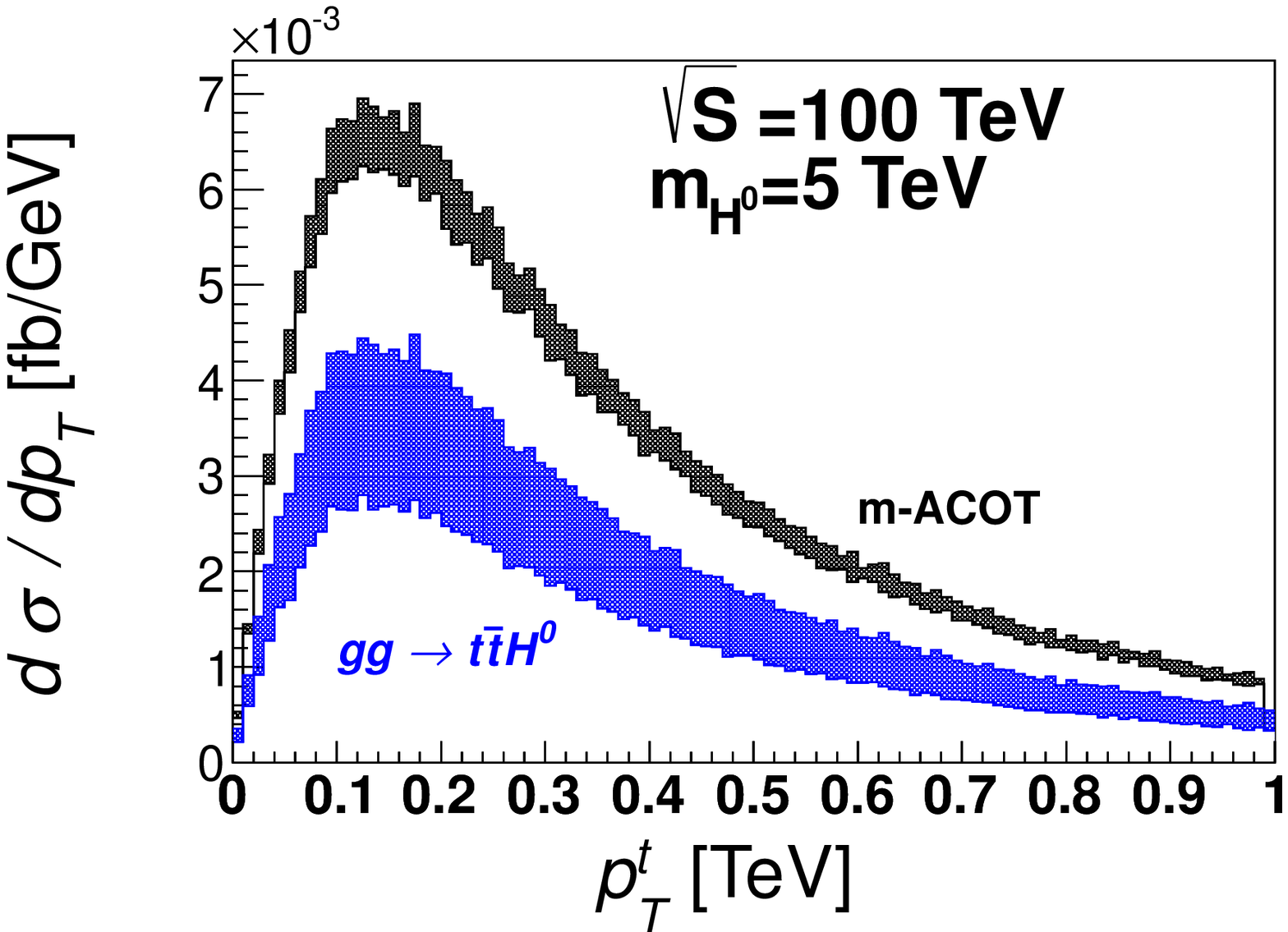} &
\hspace*{-0.2cm}
\raisebox{4.5cm}{(c)}\includegraphics[width=7.7cm]{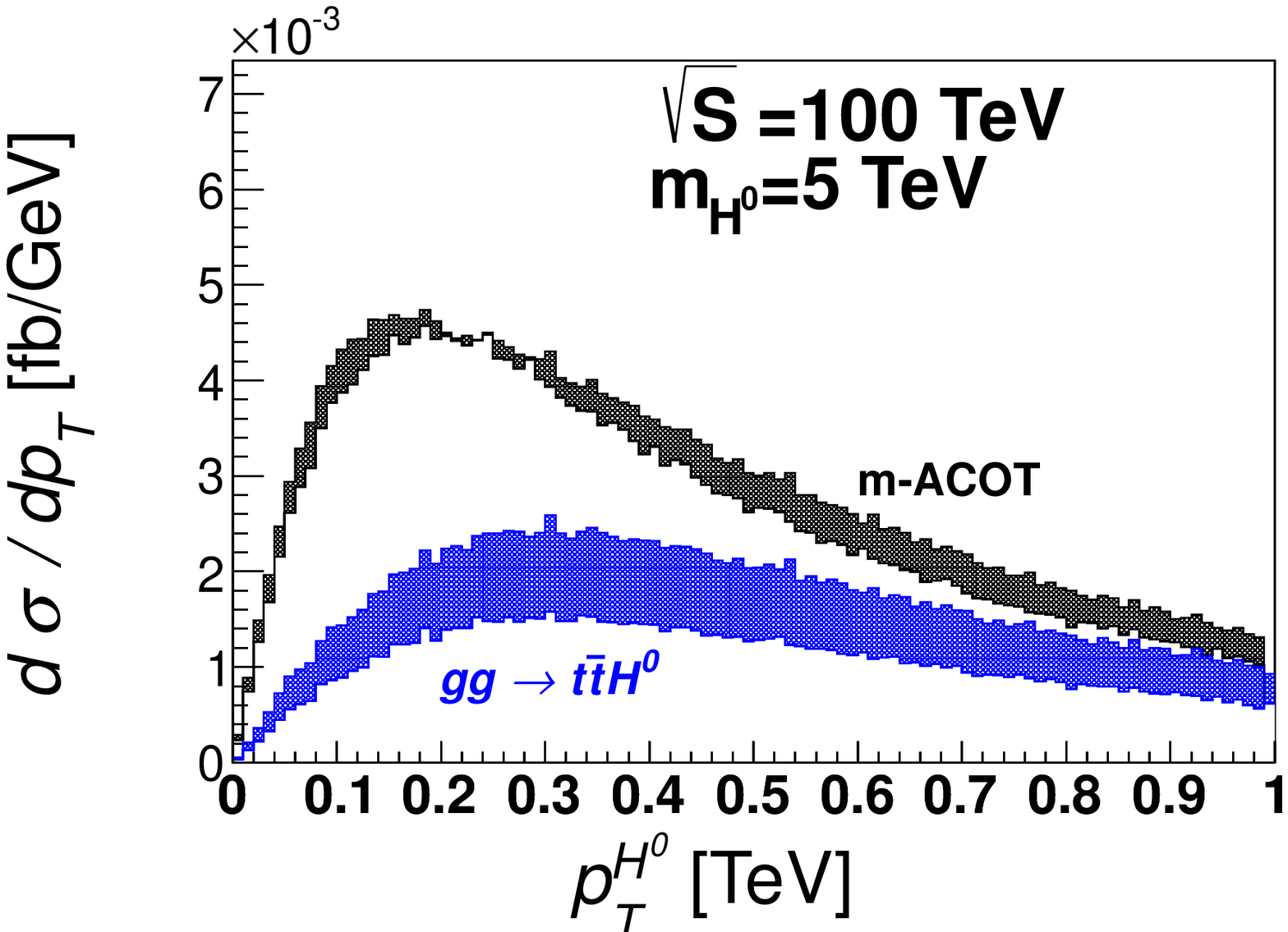}
\end{tabular}
\caption{Kinematical distributions for LO $\h$ production with $m_{\h}=5\tev$ at $\sqrt{S}=100\tev$ in m-ACOT (black) and in the 5-flavor scheme (blue). Shown are distributions of (a) the scalar-top invariant mass, $d\sigma_{pp\to \h}/dM_{\h t}$; (b) the transverse momentum of the top-quark, $d\sigma_{pp\to \h}/dp_T^t$; (c) the transverse momentum of the neutral scalar, $d\sigma_{pp\to \h}/dp_T^{\h}$. Bands indicate the scale variation in the interval $\mu_F=\mu_R\in [m_{\h}/2,2m_{\h}]$.}
\label{fig:dist}
\end{figure}
In Figure~\ref{fig:dist}, we show these three observables for $m_{\h}=5\tev$ and $\sqrt{S}=100\tev$ in the m-ACOT scheme (black) and for comparison in the 5-flavor scheme (blue). Bands indicate the scale dependence when varying $\mu_F=\mu_R$ in the interval $[m_{\h}/2,2m_{\h}]$. The reduction of the scale dependence in m-ACOT with respect to the 5-flavor prediction, as previously discussed for the total cross section in Section~\ref{sec:scale}, largely applies to the distributions as well.

The difference between the m-ACOT and 5-flavor predictions is due to higher-order collinear resummation effects. These effects are sizeable and enhance the 5-flavor result by roughly a factor of two at the maxima of the spectra. The shapes of the $M_{\h t}$ and $p_T^t$ distributions are barely affected by resummation effects, as seen in Figures~\ref{fig:dist}(a) and \ref{fig:dist}(b). The largest effect is a shift in the transverse momentum of the heavy scalar to lower $p_T^{\h}$, as seen in Figure~\ref{fig:dist}(c). This shift occurs because the average transverse momentum of the heavy scalar in the process $gg\to t\bar t\h$ is larger than that of the top-quark. The shape of the $p_T^{\h}$ distribution is also most sensitive to scale choice. For scalars heavier (lighter) than our $5\tev$ example, resummation effects in these kinematical distributions increase (decrease) with respect to Figure~\ref{fig:dist}, as expected from the behavior of the total cross section in Figure~\ref{fig:xs-dyn}, right panel.

We have also verified that the distributions are very similar for the pseudoscalar $A^0$ in Eq.~(\ref{eq:coupling}), unless its mass is close to the top threshold. In this region, small shape differences occur in the transverse momentum of the scalar versus pseudoscalar and the overall production rate of the pseudoscalar is enhanced.


\section{Discussion}\label{sec:dis}
The widely adopted s-ACOT factorization scheme is applicable to DIS and similar processes with one heavy quark in the initial state. In hadronic collisions with two initial quarks, however, the double-collinear subtraction makes the original s-ACOT approach ambiguous to implement, as discussed in detail in Section~\ref{sec:massless}. This is mended by the m-ACOT scheme proposed in this article (see Section~\ref{sec:macot}). Early work on $t\bar{t}$ and $b\bar{b}$ fusion was based on the ACOT formalism \cite{Dicus:1988cx}, although the top-quark mass was unknown at that time. Later work by the same authors used a massless scheme with heavy-quark masses retained only in the gluon-initiated process and its subtraction terms \cite{Dicus:1998hs}. This approach fails for the top-quark near the threshold, as represented in Figure~\ref{fig:zerom}. Our work updates and expands on those early results and adds a simplification in the treatment of heavy-quark masses that preserves the numerical validity of ACOT at all scales, while avoiding the pitfalls of naively applying s-ACOT.

A process closely related to the one we focus on is the $t\bar b$-initiated production of a heavy charged particle $H^+$. Our m-ACOT formalism for charged-scalar production in Eq.~(\ref{eq:xs-sacot-charged}) provides a general prescription for the consistent treatment of two different heavy-quark masses in this case. The original s-ACOT prescription is implementable by setting $m_b=0$ since then there is at most one massive quark in the initial state for $m_t \gg m_b$. Early work for $t\bar b$-initiated charged-scalar production was done in \cite{Barnett:1987jw,Olness:1987ep}. Recently, the authors of \cite{Dawson:2014pea} have investigated charged-Higgs boson production at the 100-TeV VLHC. They neglect $m_b$ and retain $m_t$ in $\bar b g \to \bar t H^+$, while neglecting $m_t$ and retaining $m_b$ in $t g \to b H^+$. This follows the spirit of the s-ACOT convention and is fully valid at high energies.
Their treatment differs from ours in that they choose to neglect the last line for $gg\to b\bar t H^+$ in Eq.~(\ref{eq:xs-sacot-charged}). In the region $m_{H^+}\gtrsim m_t$ not far above the top threshold, we expect the process $gg\to b\bar t H^+$ to be of similar size as the process $tg\to bH^+$. They also adopt the NLO PDFs and include NLO corrections of virtual and real gluons to the 6-flavor process $t\bar{b} \to H^+$, which in our framework would be accompanied by additional terms.

Although a detailed calculation at NLO \cite{Dicus:1998hs,Campbell:2002zm} in the region $m_{\h} \gtrsim m_t$ is beyond the scope of this work, it is worth considering how one would proceed for future improvement. Returning to neutral scalar production, the partonic cross section $\hat{\sigma}_{t\bar t\rightarrow \h}$ receives NLO corrections from real and virtual gluons. Convoluted with the top PDFs $f_t$, these yield terms of $\mathcal{O}(\alpha_s^3 L^2)$. When including such next-to-leading logarithm (NLL) contributions, we should also consistently include NLO corrections to $\hat{\sigma}_{tg\rightarrow t\h}$ and $\hat{\sigma}_{gg\rightarrow t\bar t\h}$ , along with subtraction terms, which are of the same order. Consistency then requires that we use NLO PDFs, since they also produce NLL terms of $\mathcal{O}(\alpha_s^3 L^2)$ when convoluted with the LO cross section $\hat{\sigma}_{t\bar t\rightarrow \h}$. In general, if we calculate QCD corrections to a given order for $t \bar t$-initiated processes, we should also include corrections at the same order to $tg/\bar t g$- and $g g$-initiated graphs to consistently account for sub-leading logarithm contributions. 

At NLO the effective PDF multiplying the LO graphs involves additional next-to-leading-logarithm subtraction terms, which will give a better approximation to $f_t$ and may not have the runaway logarithmic behavior of $f_t^0$ in the high-energy limit. This might seem to undermine our arguments for returning to the standard QCD order counting for light quarks in this limit. However, similar concerns will arise from effective PDFs with only LL subtraction terms convoluted with NLO graphs. See Appendix~\ref{sec:powersapp} for further details. 

On the numerical significance of our findings, we note a caveat: Near the top threshold the subtraction of the LL term of the top PDF in favor of a gluon splitting in the matrix element is a large negative correction. Higher-order collinear splittings are approximated by the effective PDF $f_t-f_t^0$. It may be that these higher-order terms are also overestimated compared to the explicit graphs they correspond with. One way to address this issue is with an effective scale choice as in Section~\ref{sec:scale} where a 
lower factorization scale provides better matching at the LO. In Figure~\ref{fig:scale-variation} one can see that, if we 
keep the renormalization scale fixed, this may give us a $10-20 \%$ reduction compared to the naive choice $\mu_F = m_{\h}$ at a 100-TeV collider.
Similar considerations apply at the LHC.



\section{Summary}\label{sec:sum}
When marching into the new energy frontier, the top-quark, as the heaviest particle in the SM, may play an increasingly important role in probing new physics and deepening our understanding of the SM. In this paper, we have considered top-quark initiated processes with an energy scale $Q$ above the top mass $m_t$, where the collinear enhancement factors $\alpha_s^n\log^n(Q^2/m_t^2)$ can be resummed into the top-quark PDF.
This technique may be applied to the production of heavy particles of spin-0, 1 and 2, with differences arising primarily from the matrix elements near the top threshold, as given in Table~\ref{tab:sqme}. We have illustrated our results for the scalar case. The ACOT factorization scheme with massive partons gives stable and accurate results, if all partonic processes of the same order are appropriately included (see Figure~\ref{fig:schemes1}). Neglecting heavy-quark masses in the matrix elements is appealing to simplify higher-order calculations. However, the simplified scheme (s-ACOT), developed for DIS, may lead to unphysical results near the top threshold in hadronic collisions with two initial heavy quarks, where double collinear regions occur. We propose a modified scheme, m-ACOT, that preserves some simplification while remaining valid at all scales $Q \gtrsim m_t$. Our results are summarized as follows:\\
\noindent
(1.) At the 100-TeV VLHC (14-TeV LHC), the top-quark luminosity in the relevant energy region $200\gev < \sqrt s  < 10\tev\ (4\tev)$ can be of the order of $10^4 - 0.01\ (1-10^{-5})$, which is about an order of magnitude lower than the bottom-quark luminosity (Figures~\ref{fig:lumi} and \ref{fig:pdfx}).\\
\noindent
(2.) Collinear resummation effects can enhance the inclusive cross section for the production of a heavy particle by a factor of two or more over the open-flavor associated production $gg\to t\bar t H$ for $m_H \gtrsim 5\tev$ ($0.5\tev$) at the 100-TeV VLHC (14-TeV LHC) (Figure~\ref{fig:schemes1}). Gluon-gluon fusion via a top loop dominates the inclusive production of a scalar at the LHC, but the tree-level top-quark initiated processes including resummation become larger for very high masses (Figure~\ref{fig:lhc-fcc}).\\
\noindent
(3.) The s-ACOT prescription is ambiguous in hadronic collisions with two initial heavy quarks. Applied naively, one finds unphysical results for $m_{\h}\lesssim 3\tev$ in our example process (Figure~\ref{fig:zerom}). This can be improved as in Eq.~(\ref{eq:xs-sacot-adjusted}) to 
give stable results down to the top mass threshold. However non-physical contributions remain with noticeable effects for $m_{\h} \lesssim 1\tev$. We propose to neglect the top-quark mass
 only in graphs with two initial tops in order to have a valid framework at all scales $Q \gtrsim m_t$ (m-ACOT, Section~\ref{sec:macot}).\\
\noindent
(4.) Near the top-quark threshold, a reliable calculation requires a consistent treatment of numerically comparable terms, accomplished by counting the top-quark PDF as $\alpha_s$-suppressed compared to the gluon PDF. At very high scales we recover the `light-quark' counting, which treats the top and gluon PDFs as of the same order. The top may also be treated as truly massless in this limit. In practice, we find that the massless-top limit is only reached for $Q \sim {\cal O}(10\tev)$ at a 100-TeV collider (left panel of 
Figure \ref{fig:schemes1} and Figure~\ref{fig:zerom}). \\
\noindent
(5.) The large difference between the PDF approach $t\bar t \to H$ and the open-flavor approach $gg\to t\bar t H$ can largely be attributed to the choice of the factorization scale, $\mu=m_H$. We employ an ``effective scale'', which provides a better matching of these two calculations (Figure~\ref{fig:xs-dyn}). As previous authors have found, this scale can be far below the mass of the produced particle.\\
\noindent
(6.) In general, the (m-)ACOT calculation with a consistent inclusion of LO terms is fairly stable with respect to scale variation. This is true for inclusive cross sections (Figure~\ref{fig:scale-variation}) and differential distributions (Figure~\ref{fig:dist}).\\
\noindent
Further investigation of improved predictions for top-quark initited processes is under way, which may include a full NLO calculation and NLL resummation.


\section{Acknowledgments}
We would like to thank John Collins and C.-P. Yuan for detailed comments on the manuscript.
We thank Fred Olness, Juan Rojo, and Scott Willenbrock for discussions. This work was supported in part by the U.S.~Department of Energy under Grant No.~DE-FG02-95ER40896, in part by the National Science Foundation under Grant PHY-1212635, and in part by the PITT PACC.


\appendix

\section{Factorization with heavy quarks at high energies}\label{app}
In this appendix, we present a detailed discussion of the ACOT factorization schemes for heavy quarks and their application to top-quark observables at ultra-high energies. The ACOT family in the class of General Mass Variable Flavor Number (GMVFN) schemes was developed based on 
factorization between a hard-scattering matrix element and an effective PDF which may include soft 
physics. The basis of our discussion has been developed in the framework of DIS. We adapt the notation from \cite{Collins:1998rz}. In general the structure function $F$ in DIS can be expanded as 
\begin{align}
F=C \bullet \sum\limits_{n=0}^\infty K^n \bullet T + D,
\label{eq:ladder}
\end{align}
where $C$, $K$, and $T$ represent 2-particle irreducible (2PI) graphs, which are joined by two propagator lines with an 
on-shell final-state cut running between them~\cite{Curci:1980uw}. The symbol $\bullet$ denotes the integral convolution over the momentum of the propagator connecting the different 2PI graphs, and implies a summation over flavor, color and spin indices. In the following $K$ and $T$ are defined to include the propagators connecting them to the next term on the left of Eq.~(\ref{eq:ladder}). $C$ is the graph which directly involves the probing photon of virtuality $Q^2\gg \Lambda_{\rm{QCD}}^2$ and
 $T$ includes the initial 2PI scattering from an incoming target. $D$ is a possible remaining 2PI graph which cannot be 
so decomposed. The structure function may be conveniently rewritten as \cite{Collins:1998rz}
\begin{align}
F =C \bullet \sum\limits_{n=0}^\infty [(1-Z)K]^n \bullet Z \bullet \sum\limits_{n=0}^\infty K^n \bullet T + R.
\label{eq:acot}
\end{align}
Here $R$ is a remainder term which contains corrections of $\mathcal{O}(\Lambda_{\rm{QCD}}^2/Q^2)$. $Z$ is an operator inserted between the graphs, which effectively sets the propagators connecting 2PI graphs to be collinear with the momentum of the incoming target lines.

The utility of this expansion is that we may identify $A \equiv  Z \bullet \sum_{n=0}^\infty K^n \bullet T$ as the PDF
which can be measured by experiment or extrapolated from lower scales via the DGLAP evolution equations \cite{Altarelli:1977zs}. The term $H \equiv C \bullet \sum_{n=0}^\infty [(1-Z)K]^n \bullet Z$ is the hard-scattering graph which can be calculated perturbatively.

There is some freedom in defining the operator $Z$ which preserves the fundamental properties of the factorization 
proof. In the original ACOT formulation $Z$ sets the particle lines on which it acts on-shell, meaning 
that quark masses $m_q$ are included everywhere in the calculation of $H$. However, since heavy-quark lines ultimately 
derive from gluon splitting when the original target is a proton, the propagators on which $Z$ acts have spacelike 
invariant mass. Setting them on-shell introduces an error of $\mathcal{O}(m_q^2)$ on the virtuality of the propagator, but preserves the factorization of the cross section up to corrections of $\mathcal{O}(\Lambda_{\rm{QCD}}^2/Q^2)$ \cite{Collins:1998rz}.
 A modification was proposed, known as s-ACOT, in which $Z$ acting on a quark line sets its mass to zero and all masses of the same scale or less may similarly be neglected in parts of the graph to the left of the $Z$ insertion in the expansion of Eq.~(\ref{eq:acot}). The error in doing so is of the same order as that introduced by putting the initial quark lines on-shell. In DIS all fermion lines may take the massless limit if they are nearer to the interaction with the incoming photon than the point where the $Z$ operation is applied and if their on-shell mass is less than or equal to the on-shell mass of the line on which $Z$ acts. In particular, in any graph which involves a heavy quark as an initial parton, that quark and any lighter quarks may be treated as massless throughout. 

\subsection{Power counting}\label{sec:powersapp}
It is important to note the role of the subtraction terms involving explicit $Z$ insertion in the expansion of $H$ described above. At the first order in the ACOT expansion the hard-scattering graph is simply given by
\begin{align}
H \simeq C\bullet Z.
\end{align}
To lowest order in $\alpha_s$, $C$ is given by the sum of tree-level processes $q_i \gamma \to q_i$, where $i$ runs over all quarks with mass of order $Q^2$ or less.  Here we focus on the scenario $Q^2>m_q^2$, where only one quark $q$ is massive and all other quarks with masses $m_{q}^2\ll Q^2$ can be considered massless. Light quarks will not play a role in our following discussion and are thus omitted. At the NLO in the expansion in Eq.~(\ref{eq:acot}) we have
\begin{align}\label{eq:hard}
H \simeq C\bullet Z + C \bullet (1-Z) \bullet K\bullet Z = C\bullet Z + C \bullet K\bullet Z - C \bullet Z \bullet K\bullet Z.
\end{align}
The term $C \bullet K\bullet Z$ generates the partonic process $\gamma g \to q \overline{q}$ which, convoluted with the appropriate PDFs, may be smaller than the lowest-order term $C\bullet Z$. However, the term $- C \bullet Z \bullet K\bullet Z$ is not. This negative term is usually taken to be equivalent to $C \bullet f_q^0$, where $f_q^0$ arises from the gluon splitting into a massive quark-antiquark pair, as defined for the top-quark in Eq.~(\ref{eq:top-pdf-ll}). Then the cross section at this order may be written as
\begin{align}
\int\text{d}\Pi\,\text{d}x_{1,2}\,([f_q -f_q^0]\times\hat{\sigma}_{q\gamma \rightarrow q} + f_g\times \hat{\sigma}_{g\gamma \rightarrow q\bar{q}}),
\label{eq:oneside}
\end{align} 
 where the integral over the phase space $\Pi$ and parton momentum fractions $x_{1,2}$ is implicit. 
This form has an intuitive interpretation: the second term gives the full calculation for LO gluon splitting to $q\overline{q}$, including all mass effects and 
the full phase space; the first term $\sim f_q$ gives an 
approximation to the collinear contributions at all orders resummed in the PDF $f_q$. However, the negative term  $\sim f_q^0$ subtracts out the leading logarithm of $f_q$, because this is an approximation to the collinear region in the second term. 

For scales $Q^2\gtrsim m_q^2$ not too far above the quark mass, $f_q^0$ is a good approximation to $f_q$. In this regime, one must include the subtraction term $\sim f_q^0$ along with the LO term in the ACOT expansion. Moreover, the difference $\sim [f_q-f_q^0]$ can be small and numerically sub-leading compared 
to $g \gamma \to q \overline{q}$. Numerically, all three pieces should be included at LO. This can be done by 
regarding the heavy-quark PDF as intrinsically suppressed by $\mathcal{O}(\alpha_s)$ compared to the gluon PDF, a fact which can be seen explicitly in the  LL approximation to $f_q^0$ in Eq.~(\ref{eq:top-pdf-ll}). Thus all the terms presented in Eq.~(\ref{eq:oneside}) are counted as of $\mathcal{O}(\alpha_s)$. The term  $C \bullet (1-Z) \bullet K\bullet Z$ also contains the graphs $q \gamma \to q g$ 
along with the corresponding subtraction term, while $C\bullet Z$ expanded beyond LO contains the one-loop graph for $q \gamma \to q$ involving virtual gluons. These 
pieces are of $\mathcal{O}(\alpha_s^2)$ and so are rightly counted as NLO corrections to the LO terms in Eq.~(\ref{eq:oneside}).

In the limit $Q^2/m_q^2\rightarrow \infty$, this power counting is no longer useful. In this case  the LL approximation $f_q^0$ grows indefinitely as $L=\log(Q^2/m_q^2)$, since it is generated uniquely from gluon splitting and the gluon PDF does not decrease at high $Q^2$. On the other hand, the resummed PDF $f_q$ does not grow indefinitely in a logarithmic fashion. Both 
$f_q(Q^2,x)$ and $f_q^0(Q^2,x)$ are taken to be zero at $Q^2=m_q^2$. Initially $f_q$ grows faster with $Q^2$ than $f_q^0$ due to the inclusion of resummed higher-order contributions of $\mathcal{O}(\alpha_s^nL^n)$ ($n > 1$), and quark-quark and gluon-gluon splitting in the DGLAP evolution. However, as $Q^2$ continues to increase above $m_q^2$, $f_q$ must approach an equilibrium with the gluon and lighter-quark PDFs. Thus the higher-order contributions in $f_q$ become negative and comparable to $f_q^0$. At some scale, $f_q^0$ exceeds $f_q$ and eventually, for $Q^2/m_q^2\rightarrow \infty$, $f_q^0 \gg f_q$. In the latter case it is reasonable to revise our counting of orders. Although $f_q^0$, occurring at NLO in the ACOT expansion, is larger than $f_q$ it should be comparable and opposite in sign to the positive NLO term,
\begin{align}\label{eq:sub}
\int \text{d}\Pi\,\text{d}x_{1,2}\,(f_q^0\times\hat{\sigma}_{q\gamma\rightarrow q}) \simeq \int \text{d}\Pi\,\text{d}x_{1,2}\,(f_g\times \hat{\sigma}_{g \gamma \to q \overline{q}}).
\end{align}
This is because the cross section for $g \gamma \to q \overline{q}$ on the right side is dominated by the collinear region, which is well approximated by the LL cross section for $q\gamma\rightarrow q$ on the left side. Since the difference between these two large terms is relatively small, they can be considered together as a correction to the LO cross section $q \gamma \to q$. In the limit $Q^2/m_q^2\rightarrow \infty$, the cross section at NLO is therefore given by
\begin{align}
 \int \text{d}\Pi\,\text{d}x_{1,2}\,\big(f_q\times\hat{\sigma}_{q\gamma\rightarrow q} + [f_g\times \hat{\sigma}_{g \gamma \to q \overline{q}} - f_{q}^0\times \hat{\sigma}_{q\gamma\rightarrow q}]\big).
\end{align}
With the high-energy ordering the term involving $f_q$ is the leading contribution, and the sums  involving $f_q^0$ and the gluon-initiated term it approximates give a correction suppressed by $L^{-1}$. If these corrections are included one can also consider $\mathcal{O}(\alpha_s)$ corrections to the leading heavy-quark initiated terms.  This is essentially the counting scheme advocated in \cite{Dicus:1998hs}. Similar $\mathcal{O}(\alpha_s)$ corrections to $g \gamma \to q \overline{q}$ can be considered as one order higher for the same reason that 
$g \gamma \to q \overline{q}$ and its subtraction term are treated as an NLO correction to the LO term $q \gamma \to q$. Furthermore, since in this limit $\alpha_s L$ becomes $\sim 1$ or larger,  it is useful to regard $f_q$ as no longer $\alpha_s$-suppressed compared to $f_g$.  We can treat them as the same order in an $\alpha_s$ expansion and drop the separate counting for logarithms. Therefore, in the high-energy limit we recover the order counting 
of graphs which is used for light quarks and gluons. Clearly this is also the relevant limit if we treat the heavy quark as massless throughout.

If we perform an NLO calculation near the heavy-quark scale, we will find that a similar situation arises. Again counting $f_q$ as $\alpha_s$-suppressed compared to $f_g$,  we must expand $C$ and $K$ beyond 
LO in $\alpha_s$. Thus $C\bullet Z$ will include the one-loop QCD correction to the vertex and outgoing quark in $\gamma q \to q$, and the box diagram for $\gamma g \to q \bar q$, while $C \bullet (1-Z) \bullet K\bullet Z$ will include the remaining one-loop corrections to $\gamma g \to q \bar q$, as well as the tree-level diagrams $\gamma q \to q g$. As before, we will have subtraction terms where the $Z$ operator is inserted. We must also include the next order in the ACOT expansion, $C\bullet (1-Z)\bullet K \bullet (1-Z) \bullet K\bullet Z$, although only the leading graphs in $C$ and $K$ from this term are needed, namely,  $\gamma g \to q \bar q g$. The subtraction terms for this piece will include single $Z$ insertions as well as a (positive) term with two $Z$ insertions. Schematically we can organize the full NLO calculation as
\begin{align} \label{eq:nloacot}
\int\text{d}\Pi\,\text{d}x_{1,2}\, & (f_q - f_q^1) \times \hat\sigma_{\gamma q \to q}^0 \\ \nonumber
&\hspace*{-0.5cm}+ (f_q-f_q^0) \times (\hat\sigma_{\gamma q \to q}^1 + \hat \sigma_{\gamma q \to q g}^0) + (f_g-f_g^0) \times \hat \sigma_{\gamma g \to q \bar q}^0 \\ \nonumber
&\hspace*{-0.5cm}+ f_g \times (\hat \sigma_{\gamma g \to q \bar q g}^0 + \hat \sigma_{\gamma g \to q \bar q}^1).
\end{align}
Here superscripts $0$ and $1$ indicate LO and NLO contributions to the matrix elements $\hat \sigma$ and $f_g^0$ is the LL term in the gluon PDF $f_g$, which arises from $ g \to gg$. The term $f_q^1$ contains single and double collinear splittings which correspond with LO and NLO terms in an expansion of the heavy-quark PDF. As in the LO case, $f_q^1$ subtracts out terms which are explicitly calculated from initial gluons at this order. Since $f_q^1$ contains higher order logarithm terms it does not necessarily have the runaway logarithmic behavior $f_q^0$ exhibits at large $Q^2$, and should provide a better approximation to $f_q$. This could spoil the 
argument for a change in order counting in the high-energy limit, with respect to the LO matrix element. However, the second line of Eq.~(\ref{eq:nloacot}) demonstrates that the problem of $f_q - f_q^0$  becoming large and negative arises again, where it multiplies NLO matrix elements. It is easy to see that this issue will arise at all orders, albeit suppressed by increasing powers of $\alpha_s$, so our reasoning about order 
counting holds at higher orders.

\subsection{The role of heavy quark masses}\label{sec:massesapp}
We turn now to the process of neutral scalar production from top-quark fusion. The total cross section for $pp\rightarrow t\bar t H^0$ in the ACOT scheme has been defined in Eq.~(\ref{eq:xs-acot}), where the top-quark mass is retained throughout in the partonic cross section $\hat{\sigma}(s,m_{\h},m_t)$. In analogy with the hard-scattering kernel $H$ in DIS from Eq.~(\ref{eq:hard}), $\hat{\sigma}_{t\bar t\rightarrow \h}$ corresponds to the LO kernel $C\bullet Z$, $\hat{\sigma}_{tg\rightarrow t\h}$ is the NLO contribution $C\bullet K\bullet Z$, and $\hat{\sigma}_{t\bar t\rightarrow \h}\times f_{\bar t}^0$ can be identified with the subtraction term $C\bullet Z\bullet K\bullet Z$.\footnote{Similarly, $f_t^0\times\hat{\sigma}_{t\bar t\rightarrow \h}$ is the subtraction term for the contribution $\hat{\sigma}_{g\bar t\rightarrow \bar{t}H^0}$.} However, unlike in DIS, some subtlety arises when we neglect the top-quark mass in the partonic cross section. Consider the partonic process $tg \to t\h$, which, along with its subtraction piece, contributes to the cross section as
\begin{align}\label{eq:2to2}
\int \text{d}\Pi\,\text{d}x_{1,2}\,\big(f_t(x_1)\times\hat{\sigma}_{tg\rightarrow t\h}\times f_g(x_2) - f_t(x_1)\times\hat{\sigma}_{t\bar t\rightarrow \h}\times f_{\bar t}^0(x_2)\big).
\end{align}
Graphically we represent this as 
\begin{center}
\begin{picture}(260,80)(0,0)

\ArrowLine(70,0)(70,40)
\ArrowLine(70,40)(110,40)
\ArrowLine(110,40)(110,70)
\DashLine(70,40)(70,70){3}
\Gluon(110,0)(110,40){3}{3}

\Text(50,40)[t]{$+$}
\Text(130,40)[t]{$-$}
\Text(260,40)[t]{$+ (t \to \overline{t}) + \text{s-channel}$.}                                                                                                                        
\ArrowLine(150,0)(150,40)
\ArrowLine(150,40)(190,40)
\ArrowLine(190,40)(190,70)
\DashLine(150,40)(150,70){3}
\Gluon(190,0)(190,40){3}{3}
\GCirc(170,40){3}{1}
\end{picture}
\end{center}
The circle indicates a $Z$ operator acting on the internal top-quark propagator, and such an operator is also implicitly acting on the lines of incoming quarks.
Naively one might take the massless limit for these contributions, since one of the incoming legs is a partonic top, and collinear divergences will cancel in the sum in Eq.~(\ref{eq:2to2}). According to this interpretation the $Z$ operator on the incoming top of the left figure sets masses to zero everywhere in the graph, including the logarithmic term $f_t^0$ arising from the $Z$ insertion shown in the second graph. But the inclusion of the partonic contribution $\hat{\sigma}_{gg\rightarrow t\bar t \h}$ introduces an additional set of graphs,
\begin{center}
\begin{picture}(390,100)(0,0)
\Gluon(20,0)(20,40){3}{3}
\Gluon(80,0)(80,40){3}{3}
\ArrowLine(20,40)(80,40)
\ArrowLine(20,80)(20,40)
\ArrowLine(80,40)(80,80)
\DashLine(50,40)(50,80){3}

\Gluon(115,0)(115,40){3}{3}
\Gluon(175,0)(175,40){3}{3}
\ArrowLine(115,40)(175,40)
\ArrowLine(115,80)(115,40)
\ArrowLine(175,40)(175,80)
\DashLine(145,40)(145,80){3}
\GCirc(130,40){3}{1}

\Gluon(215,0)(215,40){3}{3}
\Gluon(275,0)(275,40){3}{3}
\ArrowLine(215,40)(275,40)
\ArrowLine(215,80)(215,40)
\ArrowLine(275,40)(275,80)
\DashLine(245,40)(245,80){3}
\GCirc(260,40){3}{1}

\Gluon(310,0)(310,40){3}{3}
\Gluon(370,0)(370,40){3}{3}
\ArrowLine(310,40)(370,40)
\ArrowLine(310,80)(310,40)
\ArrowLine(370,40)(370,80)
\DashLine(340,40)(340,80){3}
\GCirc(325,40){3}{1}
\GCirc(355,40){3}{1}

\Text(5,40)[t]{$+$}
\Text(295,40)[t]{$+$}
\Text(195,40)[t]{$-$}
\Text(100,40)[t]{$-$}

\end{picture}
\end{center}
Adapting the formalism from DIS once more for our process, we find the contributions from those four diagrams contained in the hard kernel,\footnote{Here a $Z$ operator on the left-hand (right-hand) side of $C$ is acting to the right (left).}
\begin{align}
\label{eq:z2}
H & \simeq Z\bullet K\bullet(1-Z)\bullet C\bullet(1-Z)\bullet K\bullet Z\\\nonumber
 & = Z\bullet K\bullet C\bullet K\bullet Z - Z\bullet K\bullet Z\bullet C \bullet K\bullet Z - Z\bullet K\bullet C \bullet Z\bullet K\bullet Z\\\nonumber
 &\hspace*{9cm} + Z\bullet K\bullet Z\bullet C\bullet Z\bullet K\bullet Z.
\end{align}
 In the massless limit, the three rightmost graphs exhibit collinear divergences 
\begin{align}
\sim - 2\times\Big(\frac{1}{\epsilon^2} + \frac{1}{\epsilon}\Big) + \frac{1}{\epsilon^2},
\end{align}
which can be cancelled only if the top-mass is neglected in the first graph. We would thus recover the massless limit $m_t=0$, which does not appropriately describe the region $Q^2\gtrsim m_t^2$. One might take the somewhat ad-hoc prescription of keeping the set of $2 \to2$ graphs massless and 
keeping masses in the set of $2 \to 3$ graphs. As argued above this does not follow from the logic of the $Z$ operator expansion and, as shown in Section~\ref{sec:massless}, does not provide a reliable factorization scheme valid for all scales from $Q^2\gtrsim m_t^2$ up to $Q^2/m_t^2\rightarrow \infty$. 

To understand the nature of this problem one should consider the justification for s-ACOT in DIS. As mentioned above, setting the incoming quarks 
to be light-like (s-ACOT) is no worse an approximation than setting them on-shell (ACOT). In the case of internal propagators we must 
be more careful.  The explicit quark mass in the denominator of the propagator can be neglected \textit{if} the 
virtuality of that line is of $\mathcal{O}(m_t^2)$ or larger. Then we only neglect terms no larger than the error made by taking the incoming virtualities not space-like. In DIS we consider an incoming photon of virtuality $Q^2$ which interacts with a quark line connected to the incoming target 
by a series of $g \to q\overline{q}$ and $q \to qg$ splittings. For the dominant collinear contribution, the virtualities of the split partons are strongly-ordered,
\begin{align}
|k_1^2| > |k_2^2| > \ldots > |k_N^2|,
\end{align}
where $k_1^2$ is the virtuality of the splitting nearest to the photon interaction and $k_N^2$ is the farthest. If the 
initial line is  an on-shell quark, we will have $k_N^2= m_q^2$ (as in the original ACOT prescription). It follows that all the other propagators will be 
sufficiently far off-shell to justify dropping the quark mass. On the other hand, if the initial line is a gluon, then its 
virtuality may be less than $m_q^2$, at least in some regions of phase space, and we may not be able to safely ignore 
the masses in all propagators. Even if we take the massless 
limit for an incoming quark line , $k_N^2=0$ (equivalent to s-ACOT in DIS), the scheme works because the subtraction term from gluon splitting removes the  kinematic region where $k_N^2\ll m_q^2$ (cf. Eq.~(\ref{eq:sub})). For initial gluons, there is no such subtraction, unless we explicitly go to higher orders; hence the need to keep masses in graphs with external gluons.

Returning to our process $pp\rightarrow t\bar t \h$, the same sort of reasoning applies. Although the top-quark line on one side of the scalar may have virtuality $|k^2|>m_t^2$, neglecting the top mass in the  top-quark line on the other side  is not justified, if the latter comes from an initial gluon. Rather, both legs must have virtuality of $\mathcal{O}(m_t^2)$, which is not a given in those parts of the phase space where one or both of the collinear top-quarks are soft. We therefore state our interpretation of the consistent application of the simplifying principle, m-ACOT, to hadron-hadron scattering processes: The top mass can be neglected in general only in graphs where $Z$ operators are inserted on both sides of the outgoing scalar resonance. Following this instruction, we can neglect $m_t$ only in graphs with \textit{two} external top-quarks, e.g. in $t\bar t\rightarrow \h$ and its QCD corrections from virtual and real gluon radiation. At LO, this leads to the total cross section of $pp\rightarrow t\bar t \h$ in m-ACOT from Eq.~(\ref{eq:xs-sacot}). The only difference with respect to ACOT is the massless limit for the contribution $\hat{\sigma}_{t\bar t\rightarrow \h}(s=m_{\h}^2,m_t)\rightarrow \hat{\sigma}_{t\bar t\rightarrow \h}(s=m_{\h}^2,m_t=0)$.

The version of s-ACOT discussed in Eq.~(\ref{eq:xs-sacot-adjusted}) does not seem to follow in a simple way from the 
$Z$ operator formalism assumed here. It may be expressed by rewriting the last term of Eq.~(\ref{eq:z2}),
\begin{align}
Z\bullet K\bullet Z\bullet C\bullet Z\bullet K\bullet Z &\to Z\bullet K\bullet Z\bullet [C\bullet Z\bullet K]_{m=0}\bullet Z \\ \nonumber &\ +Z\bullet [K\bullet Z\bullet C]_{m=0} \bullet Z\bullet K\bullet Z \\ \nonumber &\ -Z \bullet K\bullet Z\bullet [C]_{m=0}\bullet Z\bullet K\bullet Z,
\end{align}
where $Z$ operators are acting differently in each term to set masses to zero inside the brackets as shown.

\section{Cross section for top-initiated scalar production}\label{sec:part-xs}
In this appendix, we provide analytic results for the partonic cross section of the top-quark initiated processes $t\bar t\to H^0$ and $tg\to H^0$ at the tree level. Analytic formulae for the gluon-initiated process $gg\to t\bar t H^0$ and the loop-induced process $gg\to H^0$ can be found in \cite{Dawson:2003zu} and \cite{Georgi:1977gs,Djouadi:2005gi}, respectively.
\paragraph{The process $t\bar t\to H^0$:} The partonic cross section at the tree level is given by
\begin{align}\label{eq:stt}
\hat{\sigma}_{t\bar t\to H^0}(s) = \frac{1}{\Phi_{t\bar t}}2\pi\delta(s-m_{H^0}^2)\overline\sum |{\cal M}_{t\bar  t\to H^0}|^{2},\qquad \Phi_{t\bar t}=2s\beta_{t\bar t},
\end{align}
with $\overline\sum |{\cal M}_{t\bar  t\to H^0}|^{2}$ and $\beta_{t\bar t}$ defined in Table~\ref{tab:sqme}. The contribution to the hadronic cross section $\sigma_{pp\to H^0}$ is obtained by folding $\hat{\sigma}_{t\bar t\to H^0}(s)$ with the top-quark PDFs according to Eq.~(\ref{eq:factorization}).
\paragraph{The process $tg\to tH^0$:} We assign the parton momenta as $t(p_t)g(p_g)\to t(p_{t'})H^0(p_H)$ and define the Mandelstam variables $s=(p_t+p_g)^2$ and $t = (p_{t'}-p_g)^2$. At the tree level, the differential partonic cross section differential is given by
\begin{align}\label{eq:dsdt}
\frac{d\hat{\sigma}_{tg\to H^0}}{dt} =\frac{1}{\Phi_{tg}}\frac{1}{16\pi}\,\overline\sum |{\cal M}_{tg\to tH^0}|^{2},\qquad \Phi_{tg} = 2s\beta_{tg},
\end{align}
with the spin- and color-averaged squared matrix element
\begin{align}\label{eq:metg}
\overline\sum |{\cal M}_{tg\to tH^0}|^{2} & = \alpha_s\,y^2\frac{\pi}{3}\Bigg\{2\,\frac{-st+m_{H^0}^2(s+t+4m_t^2)-m_{H^0}^4-m_t^2(s+t+5m_t^2)}{(t-m_t^2)(s-m_t^2)}\\\nonumber
& + \frac{-st+m_t^2(s-3t+2m_{H^0}^2 - 5m_t^2)}{(t-m_t^2)^2} + \frac{-st + m_t^2(t-3s+2m_{H^0}^2 - 5m_t^2)}{(s-m_t^2)^2}\Bigg\}.
\end{align}
The total partonic cross section is readily obtained by integrating Eq.~(\ref{eq:dsdt}) over $t$,
\begin{align}\label{eq:stg}
\hat{\sigma}_{tg\to H^0}(s) = \int_{t^-}^{t^+}dt\,\frac{d\hat{\sigma}_{tg\to H^0}}{dt},
\end{align}
within the integration boundaries
\begin{align}\label{eq:tpm}
 t^{\pm} & = m_{H^0}^2+m_t^2 -\frac{s}{2}\Bigg\{\Big(1+\frac{m_t^2}{s}\Big)\Big(1+\frac{m_{H^0}^2-m_t^2}{s}\Big)\\\nonumber
& \qquad\qquad\qquad\qquad\qquad\qquad\pm\Big(1-\frac{m_t^2}{s}\Big)\Bigg[\Big(1-\frac{m_{H^0}^2-m_t^2}{s}\Big)^2-\frac{4m_t^2}{s}\Bigg]^{1/2}\Bigg\}.
\end{align}
The expression in Eq.~(\ref{eq:stg}) also describes the partonic cross section $\hat{\sigma}_{\bar tg\to H^0}(s)$ for $\bar t g \to \bar t H^0$. The contribution to the hadronic cross section $\sigma_{pp\to H^0}$ is given by folding $\hat{\sigma}_{t(\bar t)g\to H^0}(s)$ in with the top-quark and gluon PDFs as in Eq.~(\ref{eq:factorization}).

Note that in the massless-top limit $m_t\to 0$ the lower boundary $t^-$ in Eq.~(\ref{eq:tpm}), which corresponds to collinear top-quark emission, goes to zero. The upper boundary $t^+$, corresponding to anti-collinear top emission, approaches $t^+ =  m_{H^0}^2-s$, which vanishes if the emitted top-quark is soft. The squared matrix element in Eq.~(\ref{eq:metg}) thus diverges in these phase-space regions for $m_t\to 0$, because the virtuality $t$ in the internal top-quark propagator is no longer regularized by the top mass. This leads to the problems with the massless-top approximation we discussed in Section~\ref{sec:massless}.




\begin{thebibliography}{99}
\bibitem{SnowHiggs} 
  S.~Dawson, A.~Gritsan, H.~Logan, J.~Qian, C.~Tully, R.~Van Kooten, A.~Ajaib and A.~Anastassov {\it et al.},
  arXiv:1310.8361 [hep-ex].

\bibitem{Gershtein:2013iqa} 
  Y.~Gershtein, M.~Luty, M.~Narain, L.-T.~Wang, D.~Whiteson, K.~Agashe, L.~Apanasevich and G.~Artoni {\it et al.},
  arXiv:1311.0299 [hep-ex].

\bibitem{Hook:2014rka} 
  A.~Hook and A.~Katz,
  JHEP {\bf 1409}, 175 (2014)
  [arXiv:1407.2607 [hep-ph]].

\bibitem{Barnett:1987jw} 
  R.~M.~Barnett, H.~E.~Haber and D.~E.~Soper,
  Nucl.\ Phys.\ B {\bf 306}, 697 (1988).

\bibitem{Olness:1987ep} 
  F.~I.~Olness and W.~K.~Tung,
  Nucl.\ Phys.\ B {\bf 308}, 813 (1988).

\bibitem{Abdesselam:2010pt} 
  A.~Abdesselam, E.~B.~Kuutmann, U.~Bitenc, G.~Brooijmans, J.~Butterworth, P.~Bruckman de Renstrom, D.~Buarque Franzosi and R.~Buckingham {\it et al.},
  Eur.\ Phys.\ J.\ C {\bf 71}, 1661 (2011)
  [arXiv:1012.5412 [hep-ph]].

\bibitem{Altarelli:1977zs} 
  V.~N.~Gribov and L.~N.~Lipatov,
  Sov.\ J.\ Nucl.\ Phys.\  {\bf 15}, 438 (1972)
  [Yad.\ Fiz.\  {\bf 15}, 781 (1972)];
  G.~Altarelli and G.~Parisi,
  Nucl.\ Phys.\ B {\bf 126}, 298 (1977).

\bibitem{Thorne:1997ga} 
  R.~S.~Thorne and R.~G.~Roberts,
  Phys.\ Rev.\ D {\bf 57}, 6871 (1998)
  [hep-ph/9709442].

\bibitem{Aivazis:1993pi} 
  M.~A.~G.~Aivazis, J.~C.~Collins, F.~I.~Olness and W.~K.~Tung,
  Phys.\ Rev.\ D {\bf 50}, 3102 (1994)
  [hep-ph/9312319].

\bibitem{Collins:1998rz} 
  J.~C.~Collins,
  Phys.\ Rev.\ D {\bf 58}, 094002 (1998)
  [hep-ph/9806259].

\bibitem{Kramer:2000hn} 
  M.~Kr\"amer, F.~I.~Olness and D.~E.~Soper,
  Phys.\ Rev.\ D {\bf 62}, 096007 (2000)
  [hep-ph/0003035].

\bibitem{Dicus:1988cx} 
  D.~A.~Dicus and S.~Willenbrock,
  Phys.\ Rev.\ D {\bf 39}, 751 (1989).

\bibitem{Dicus:1998hs} 
  D.~Dicus, T.~Stelzer, Z.~Sullivan and S.~Willenbrock,
  Phys.\ Rev.\ D {\bf 59}, 094016 (1999)
  [hep-ph/9811492].

\bibitem{Maltoni:2003pn} 
  F.~Maltoni, Z.~Sullivan and S.~Willenbrock,
  Phys.\ Rev.\ D {\bf 67}, 093005 (2003)
  [hep-ph/0301033].

\bibitem{Maltoni:2012pa} 
  F.~Maltoni, G.~Ridolfi and M.~Ubiali,
  JHEP {\bf 1207}, 022 (2012)
  [Erratum-ibid.\  {\bf 1304}, 095 (2013)]
  [arXiv:1203.6393 [hep-ph]].

\bibitem{Dawson:2014pea} 
  S.~Dawson, A.~Ismail and I.~Low,
  Phys.\ Rev.\ D {\bf 90}, 014005 (2014)
  [arXiv:1405.6211 [hep-ph]].

\bibitem{Dittmaier:2011ti}
   S.~Dittmaier {\it et al.}  [LHC Higgs Cross Section Working Group Collaboration],
   arXiv:1101.0593 [hep-ph].

\bibitem{Hill:2002ap}
   C.~T.~Hill and E.~H.~Simmons,
   Phys.\ Rept.\  {\bf 381}, 235 (2003)
   [Erratum-ibid.\  {\bf 390}, 553 (2004)]
   [hep-ph/0203079];
  K.~Agashe, A.~Delgado, M.~J.~May and R.~Sundrum,
  JHEP {\bf 0308}, 050 (2003)
  [hep-ph/0308036].

\bibitem{KK}
  A.~L.~Fitzpatrick, J.~Kaplan, L.~Randall and L.~T.~Wang,
  JHEP {\bf 0709}, 013 (2007)
  [hep-ph/0701150];
  H.~Davoudiasl, J.~L.~Hewett and T.~G.~Rizzo,
  Phys.\ Rev.\ D {\bf 63}, 075004 (2001)
  [hep-ph/0006041];
   T.~Han, J.~D.~Lykken and R.~J.~Zhang,
   Phys.\ Rev.\ D {\bf 59}, 105006 (1999)
   [hep-ph/9811350];
  G.~F.~Giudice, R.~Rattazzi and J.~D.~Wells,
  Nucl.\ Phys.\ B {\bf 544}, 3 (1999)
  [hep-ph/9811291].

\bibitem{Ball:2012cx} 
  R.~D.~Ball, V.~Bertone, S.~Carrazza, C.~S.~Deans, L.~Del Debbio, S.~Forte, A.~Guffanti and N.~P.~Hartland {\it et al.},
  Nucl.\ Phys.\ B {\bf 867}, 244 (2013)
  [arXiv:1207.1303 [hep-ph]].

\bibitem{Buza:1996wv} 
  M.~Buza, Y.~Matiounine, J.~Smith and W.~L.~van Neerven,
  Eur.\ Phys.\ J.\ C {\bf 1}, 301 (1998)
  [hep-ph/9612398].

\bibitem{Campbell:2002zm} 
  J.~M.~Campbell, R.~K.~Ellis, F.~Maltoni and S.~Willenbrock,
  Phys.\ Rev.\ D {\bf 67}, 095002 (2003)
  [hep-ph/0204093].

\bibitem{Collins&Yuan}
J.~Collins and C.-P.~Yuan, private communication.

\bibitem{Kretzer:2003it} 
  S.~Kretzer, H.~L.~Lai, F.~I.~Olness and W.~K.~Tung,
  Phys.\ Rev.\ D {\bf 69}, 114005 (2004)
  [hep-ph/0307022].

\bibitem{Cacciari:1998it} 
  M.~Cacciari, M.~Greco and P.~Nason,
  JHEP {\bf 9805}, 007 (1998)
  [hep-ph/9803400].

\bibitem{Forte:2010ta} 
  S.~Forte, E.~Laenen, P.~Nason and J.~Rojo,
  Nucl.\ Phys.\ B {\bf 834}, 116 (2010)
  [arXiv:1001.2312 [hep-ph]].

\bibitem{Tung:2001mv} 
  W.~K.~Tung, S.~Kretzer and C.~Schmidt,
  J.\ Phys.\ G {\bf 28}, 983 (2002)
  [hep-ph/0110247].

\bibitem{Hahn:2004fe} 
  T.~Hahn,
  Comput.\ Phys.\ Commun.\  {\bf 168}, 78 (2005)
  [hep-ph/0404043].

\bibitem{madgraph} 
  J.~Alwall, R.~Frederix, S.~Frixione, V.~Hirschi, F.~Maltoni, O.~Mattelaer, H.-S.~Shao and T.~Stelzer {\it et al.},
  JHEP {\bf 1407}, 079 (2014)
  [arXiv:1405.0301 [hep-ph]].

\bibitem{Dawson:2003zu} 
  S.~Dawson, C.~Jackson, L.~H.~Orr, L.~Reina and D.~Wackeroth,
  Phys.\ Rev.\ D {\bf 68}, 034022 (2003)
  [hep-ph/0305087].

\bibitem{Georgi:1977gs} 
  H.~M.~Georgi, S.~L.~Glashow, M.~E.~Machacek and D.~V.~Nanopoulos,
  Phys.\ Rev.\ Lett.\  {\bf 40}, 692 (1978).

\bibitem{Djouadi:2005gi} 
  A.~Djouadi,
  Phys.\ Rept.\  {\bf 457}, 1 (2008)
  [hep-ph/0503172].

\bibitem{Curci:1980uw} 
  G.~Curci, W.~Furmanski and R.~Petronzio,
  Nucl.\ Phys.\ B {\bf 175}, 27 (1980).

\end{thebibliography}
\end{document}